\documentclass[12pt,3p]{elsarticle}
\usepackage[linesnumbered,ruled,vlined]{algorithm2e}
\usepackage{algpseudocode}
\let\oldnl\nl
\newcommand{\nonl}{\renewcommand{\nl}{\let\nl\oldnl}}
\usepackage[T1]{fontenc}
\usepackage{longtable}
\usepackage{makecell}
\newcommand{\rc}[1]{\textcolor{black}{#1}} 
\usepackage{graphicx}
\usepackage{amsmath,amssymb,amsfonts}
\usepackage[table]{xcolor}
\usepackage[utf8]{inputenc}
\usepackage{textcomp}
\usepackage{float}
\usepackage{orcidlink}
\usepackage{subcaption}
\usepackage{adjustbox}
\usepackage{rotating}
\usepackage{csquotes}
\usepackage{lineno,hyperref}
\hypersetup{colorlinks,
citecolor=blue,colorlinks=true,linkcolor=blue, urlcolor=blue, linktocpage
}
\usepackage{booktabs,multirow,array}
\usepackage{tabularx}
\usepackage{ragged2e}
\usepackage{bm}
\usepackage{pgfplots}
\usepackage{tikz}
\usetikzlibrary{shadows,arrows.meta,positioning}
\usetikzlibrary{shapes.geometric, backgrounds, fit, decorations.pathreplacing, calc}
\usepackage{float} 
\usepackage{pgf-pie} 
\usepackage[outline]{contour} 
\usepgfplotslibrary{groupplots}
\usetikzlibrary{shapes.geometric, arrows, positioning}
\usepackage{enumitem}
\usepackage{xcolor}
\usepackage{pifont} 
\usepackage{pgfplotstable}
\modulolinenumbers[5]
\definecolor{orcidlogocol}{HTML}{A6CE39}

\newcommand{\orcid}[1]{\href{https://orcid.org/#1}{\textcolor{orcidlogocol}{\faOrcid~}}}
\usepackage{etoolbox}
\journal{Computer Science Reviews}

\begin{document}
\begin{frontmatter}

\title{\rc{Artificial Intelligence Driven Channel Coding and Resource Optimization for Wireless Networks: A Systematic Survey}}


\author[1,1a]{Yasir Ali\orcidlink{0000-0003-3998-9218}}
\author[4]{Tayyab Manzoor\orcidlink{0000-0003-3932-0506}} 
\author[1]{Huan Yang\orcidlink{0000-0001-8539-7104}}
\author[3]{Chenhang Yan\orcidlink{0000-0002-0885-8787}} 
\author[2,1]{Yuanqing Xia\orcidlink{0000-0002-5977-4911}\corref{mycorrespondingauthor}}
\ead{xia\_yuanqing@bit.edu.cn} 
\cortext[mycorrespondingauthor]{Corresponding Author}
\address[1]{School of Automation, Beijing Institute of Technology, Beijing 100081, China. 
}
\address[1a]{Zhengzhou Research Institute, Beijing Institute of Technology, Zhengzhou 450000, Henan Province, China. 
}
\address[4]{School of Automation and Electrical Engineering, Zhongyuan University of Technology, Zhengzhou 450007, Henan Province, China.}
\address[2]{Zhongyuan University of Technology, Zhengzhou 450007, Henan Province, China. 
}
\address[3]{Zhejiang Key Laboratory of Intelligent Perception and Control for Complex Systems, Zhejiang University of Technology, Hangzhou, China.
}

\begin{abstract}
 The ongoing evolution of 5G and its enhanced version, 5G+, has significantly transformed the telecommunications landscape, driving an unprecedented demand for ultra-high-speed data transmission, ultra-low latency, and resilient connectivity. These capabilities are essential for enabling mission-critical applications such as the Internet of Things, autonomous vehicles, and smart city infrastructures. \rc{This survey} investigates the important role of Artificial Intelligence (AI) in addressing the key challenges faced by 5G/5G+ networks, including interference mitigation, dynamic resource allocation, and maintaining seamless network operation. The study particularly focuses on AI-driven innovations in coding theory, which offer advanced solutions to the limitations of conventional error correction and modulation techniques. By employing deep learning, reinforcement learning, and neural network-based approaches, including convolutional neural networks, recurrent neural networks, and Transformer-based models, this research demonstrates significant advancements in error correction performance, decoding efficiency, and adaptive transmission strategies. Additionally, the integration of AI with emerging technologies, such as massive multiple-input and multiple-output, intelligent reflecting surfaces, and privacy-enhancing mechanisms, is discussed, highlighting their potential to propel the next generation of wireless networks. \rc{This survey} provides an insightful overview of the transformative impact of AI on modern wireless communication, establishing a foundation for scalable, adaptive, and more efficient network architectures.
\end{abstract}

\end{frontmatter}

\section{Introduction}\label{Introduction}
The rapid evaluation of telecommunications technology has introduced the era of 5G and enhanced version ``5G+'' networks, which are set to drastically reshape global communication systems \cite{bertuletti202566, panek20235g}. As the need for increased data rates, extremely low latency, and the ability to support more connected devices continues to rise, the limitations of traditional wireless communication methods are becoming increasingly clear \cite{brinton2025key, evans2024satellite, zhan2025review}. These evolving needs, driven by emerging applications like autonomous systems \cite{mahalingam2025reliable, BagheriTowardConnected2021}, augmented reality \cite{RenAnEdgeAssisted2022}, and large-scale Internet of Things (IoT) networks \cite{JiangHighEfficient2024}, etc, require a fundamental shift in how these networks are designed, managed, and optimized \cite{panek20235g}. Artificial Intelligence (AI) is a key enabler in this transition, addressing complex challenges and enhancing the efficiency of next-generation wireless systems \cite{HuangAICoding2020}.

5G and 5G+ networks offer ultra-high throughput, minimal latency, and massive device connectivity \cite{bojkovic20215g}, supporting advanced applications like autonomous vehicles, smart cities, and IoT systems \cite{Chukhno5GIoTNetworks2022}. As shown in Table \ref{tabcomp2}, their standout feature is the ability to manage vast data volumes at high speeds with low latency, distinguishing them from prior generations. Despite these strengths, several challenges must be addressed for optimal performance. Important among them is signal interference, particularly in dense environments where overlapping transmissions degrade quality \cite{zhang2024massive}. Effective interference mitigation is essential to ensure uninterrupted communication \cite{VanDeepLearningBased2022, ModelDriven2020}. Another key issue is dynamic resource allocation \cite{ly2021review}, requiring real-time distribution of bandwidth and computing power to accommodate fluctuating demands \cite{EiEfficientResourceAllocation2022}. Ensuring critical applications receive priority while maintaining overall efficiency is difficult, especially given the rapid IoT expansion. Therefore, maintaining seamless connectivity across varied environments urban, rural, indoor, and outdoor, is challenging \cite{ELRAJAB2024110294}. Devices must retain stable links while transitioning between settings. Traditional static network architectures struggle with this complexity, making them ill-suited to 5G+'s dynamic requirements \cite{LiDecAge2024}.

Emerging requirements from enhanced Mobile Broadband (eMBB), massive Machine-Type Communications (mMTC), and Ultra-Reliable Low Latency Communications (URLLC) are shaping next-generation wireless networks \cite{KhanURLLCandeMBB2022}. These paradigms support applications such as high-definition video streaming, virtual reality, and ultra-low latency services. eMBB requires high data rates enabled by mmWave and massive MIMO \cite{ruan2023simplified, paper14DeepLearning-BasedChannelEstimation2018}, while mMTC focuses on large-scale, low-power IoT connectivity using LPWAN and NOMA \cite{mahmood2021machine}. URLLC targets latency-critical applications through edge computing and adaptive resource allocation \cite{masaracchia2021uav, Aliieeeaccess2024, ali2024reliability}. Meeting these demands requires high throughput, reliability, low latency, and energy efficiency.

To address these challenges, AI and Machine Learning (ML) techniques are increasingly integrated into 5G/5G+ networks. AI enables traffic prediction, adaptive resource allocation, and interference management, improving system efficiency and scalability \cite{zhangAIfor5G&B5G2020}. By learning from network data, ML models support real-time decision-making and enhance reliability under dynamic conditions.

A key application of AI lies in channel coding, which ensures reliable data transmission under noise and interference \cite{HuangAICoding2020, liaoContrucofpolarCode2022}. Traditional coding schemes, such as Reed–Solomon and Hamming codes, struggle under complex channel conditions \cite{rani2022efficient}, while conventional decoding methods (e.g., SC, BP, and Viterbi) face scalability and complexity limitations \cite{bioglio2020design, LiangNeuralEnhancedBelief2021, wu2024micro}. AI-driven approaches, including deep learning (DL)-based coding, adaptive modulation, and joint source-channel coding \cite{liaoContrucofpolarCode2022,liu2020deep, wang2020adaptive, WuDeepJointSource2024}, provide improved error correction, efficiency, and adaptability. These methods enhance transmission reliability and robustness under varying network conditions \cite{tian2024IntelligentSystemofCollege2023}.

AI integration in 5G+ extends beyond coding to network management, including resource allocation, scheduling, and QoS optimization. AI-driven solutions enable automated network configuration, traffic optimization, and failure prediction, improving overall system resilience and efficiency \cite{zhangAIfor5G&B5G2020}.

To this end, \rc{this survey seeks} to provide researchers with clear insights into the applications of AI in networking, channel coding, and security within communication systems, prioritizing conceptual understanding and clarity over extensive mathematical detail. The primary contributions of this survey are summarized as follows:

\begin{itemize}[label=--]
    \item We provide a comprehensive and \rc{up-to-date survey} of AI techniques across networking, channel coding, and security for 5G/5G+ systems, covering core technical modules and new emerging paradigms.
    
    \item We offer a critical analysis and categorization of AI-based coding and decoding methods, including novel comparative tables and diagrams, clarifying their operational mechanisms, strengths, and limitations.
    
    \item We survey and evaluate the intersection of AI with security and privacy in 5G+, highlighting new vulnerabilities and advanced AI-driven mitigation approaches.
    
    \item We identify open challenges and propose a forward-looking roadmap for future research, emphasizing key gaps and opportunities for scalable, secure, and efficient AI integration in next-generation wireless networks.
\end{itemize}

To address these elements, the remainder of this paper is organized as follows. Section~\ref{Introduction} introduces the survey framework by presenting the background and related work together with the survey scope and literature selection methodology. Section~\ref{AI-DrivenDetection} reviews AI-driven detection and precoding techniques for MIMO systems. Section~\ref{AIinCodingTheory} examines AI-based advances in channel coding for 5G/5G+ networks, covering coding fundamentals, learning-based decoding methods, neural BP-enhanced decoding, AI-driven code construction, and newly added critical discussions on the comparative effectiveness, failure modes, and contradictory findings of DL-based channel coding approaches. Section~\ref{otherAI-EnhancedBaseband} discusses other AI-enhanced baseband components, including equalization, NOMA/SCMA, and channel estimation, together with broader semantic communication-related developments. Section~\ref{SecurityandPrivacy} addresses security and privacy issues in AI-driven 5G+ networks, including model vulnerabilities, privacy-preserving learning approaches, quantum communication as a long-term direction, and the advantages and limitations of AI-based security solutions. Section~\ref{FutureResearchDirections} presents future research directions, including open research challenges, practical deployment challenges, system-level trade-offs, and future directions of AI-driven learning in wireless networks. Finally, Section~\ref{Conclusion} concludes the survey.

\begin{figure*}[htbp!]
\includegraphics[width=\linewidth]{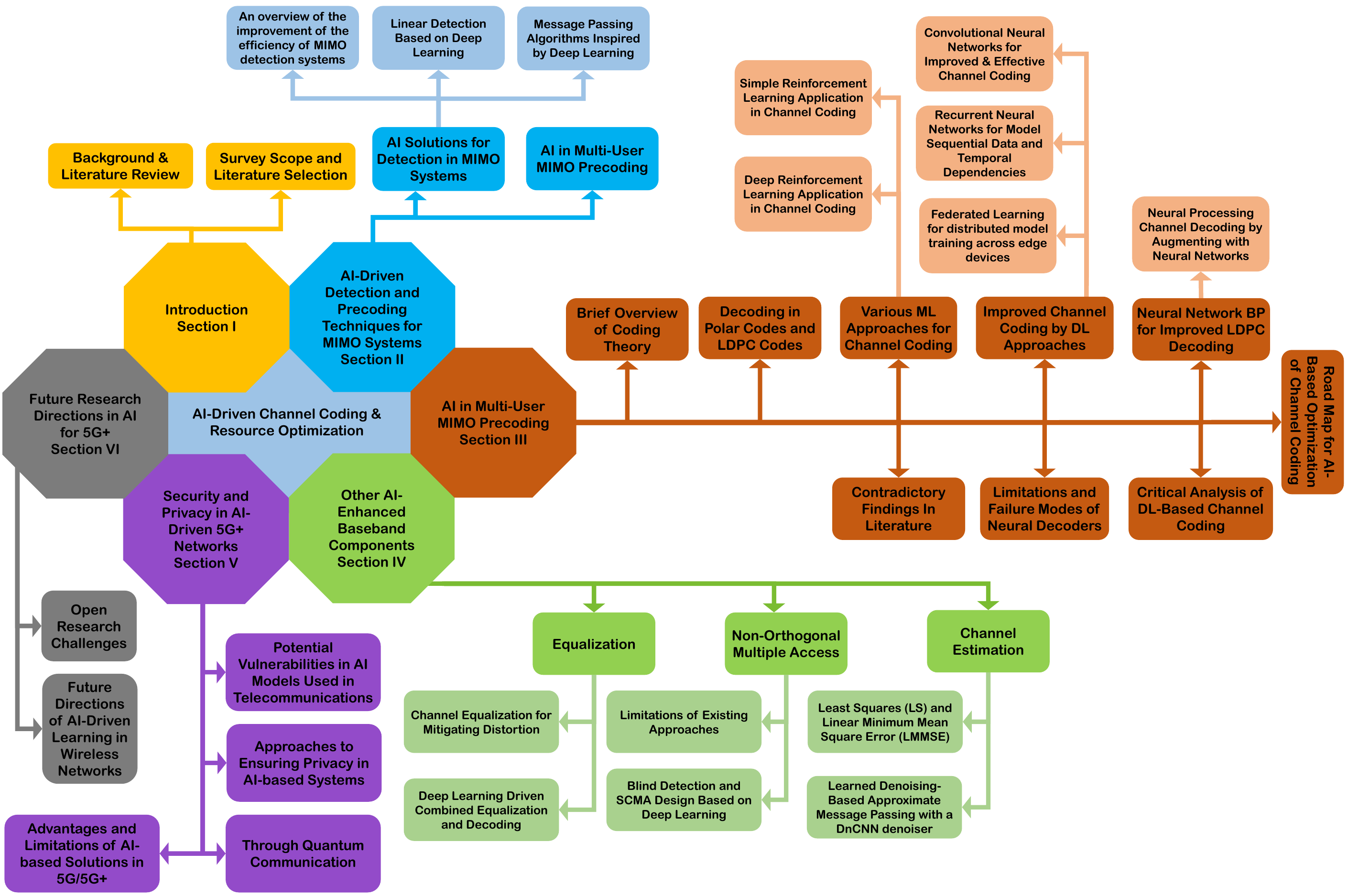}
\centering
\caption{Illustrates the organization and logical flow of this survey, highlighting the central role of AI-driven channel coding and its interaction with detection, precoding, and other baseband components.}
\label{Fig1.0}
\end{figure*}

\begin{table*}[htbp!]
\centering
\caption{Comprehensive Comparison on Key Differences Between 4G, 5G, 5G+, and 6G}
\label{tabcomp2}
\resizebox{\textwidth}{!}{%
\begin{tabular}{cccccccc}
\toprule
\begin{tabular}[c]{@{}c@{}}\textbf{Feature}\\ \textbf{Category}\end{tabular} & \textbf{4G (LTE)} & \textbf{5G (Standard)} & \textbf{5G+ (Enhanced 5G)} & \textbf{6G (Future)} \\ \midrule
\begin{tabular}[c]{@{}c@{}}\textbf{Max Speed}\\ \textbf{(Mbps)}\end{tabular} & Up to 1 Gbps (theoretical) & 10 Gbps (theoretical) & 10–20 Gbps (real-world max) & 100 Gbps–1 Tbps (theoretical) \\ \midrule
\begin{tabular}[c]{@{}c@{}}\textbf{Latency}\\ \textbf{(ms)}\end{tabular} & 30–50 ms & 1–10 ms & $<$ 1 ms & $<$ 1 ms \\ \midrule
\begin{tabular}[c]{@{}c@{}}\textbf{Device Density}\\ \textbf{(devices/km²)}\end{tabular} & 100,000 devices/km² & 1 million devices/km² & 1 million+ devices/km²  (optimized) & 1 million+ devices/km² (optimized)\\ \midrule
\begin{tabular}[c]{@{}c@{}}\textbf{Spectrum}\\ \textbf{(GHz)}\end{tabular} & Sub-6 GHz (mainly) & Sub-6 GHz + mmWave & mmWave ($>$24 GHz), Sub-6 GHz & THz, Sub-6 GHz, mmWave \\ \midrule
\begin{tabular}[c]{@{}c@{}}\textbf{Bandwidth}\\ \textbf{(MHz)}\end{tabular} & 20-100 MHz (typical) & 100 MHz–1 GHz & $>$ 1 GHz & Up to 100 GHz \\ \midrule
\begin{tabular}[c]{@{}c@{}}\textbf{Connection}\\ \textbf{Type}\end{tabular} & Mobile Broadband & \rc{eMBB}, mMTC, URLLC & \begin{tabular}[c]{@{}c@{}}eMBB+, mMTC, URLLC\end{tabular} & \begin{tabular}[c]{@{}c@{}}uMBB, mMTC, URLLC\end{tabular} \\ \midrule
\begin{tabular}[c]{@{}c@{}}\textbf{Network}\\ \textbf{Topology}\end{tabular} & Traditional cellular grid & Dense small cells & High-density mmWave cells & AI-driven, intelligent, dynamic topology \\ \midrule
\begin{tabular}[c]{@{}c@{}}\textbf{Energy}\\ \textbf{Efficiency}\end{tabular} & Moderate & Improved efficiency & High (optimized for IoT) & Ultra-high (energy-aware design) \\ \midrule
\begin{tabular}[c]{@{}c@{}}\textbf{Coverage}\\ \textbf{Area}\end{tabular} & Global coverage & \begin{tabular}[c]{@{}c@{}}Wide coverage (slightly\\ reduced vs. 4G)\end{tabular} & Limited range (due to mmWave) & \begin{tabular}[c]{@{}c@{}}Global, ubiquitous coverage \\ (via AI and mesh networks)\end{tabular} \\ 
\midrule
\begin{tabular}[c]{@{}c@{}}\textbf{Latency}\\ \textbf{Sensitivity}\end{tabular} & \begin{tabular}[c]{@{}c@{}}Moderate latency tolerance \\ (e.g., multimedia services)\\(e.g., gaming, video)\end{tabular}& \begin{tabular}[c]{@{}c@{}}Ultra-low latency (e.g., autono-\\mous vehicles, industrial IoT)\end{tabular} & \begin{tabular}[c]{@{}c@{}}Ultra-low latency (e.g.,\\AR/VR, edge computing)\end{tabular} & \begin{tabular}[c]{@{}c@{}}Ultra-low latency, real-time AI/ML \\ (e.g., brain-machine interfaces)\end{tabular} \\ \midrule
\begin{tabular}[c]{@{}c@{}}Key\\ \textbf{Use Cases}\end{tabular} & Smartphones, video streaming & \begin{tabular}[c]{@{}c@{}}High-speed internet\\AR/VR, smart cities\end{tabular} & \begin{tabular}[c]{@{}c@{}}IoT, autonomous EVs, AR/VR,\\Telemedicine 2.0, CCSs etc \end{tabular} & \begin{tabular}[c]{@{}c@{}}AI, BCIs, IoE, Autonomous Vehicles\\MR, Telemedicine 2.0, CCSs etc \end{tabular} \\ \midrule
\begin{tabular}[c]{@{}c@{}}\textbf{Peak}\\ \textbf{Throughput}\end{tabular} & 1 Gbps & 10 Gbps & 10-20 Gbps & 100 Gbps–1 Tbps \\ \midrule
\begin{tabular}[c]{@{}c@{}}\textbf{MIMO}\\ \textbf{Technology}\end{tabular} & 2x2 MIMO, 4x4 MIMO & Massive MIMO (64x64) & \begin{tabular}[c]{@{}c@{}}Advanced Massive\\ MIMO (256x256)\end{tabular}
 & \begin{tabular}[c]{@{}c@{}}Reconfigurable intelligent\\surfaces (RIS), advanced MIMO\end{tabular} \\ \midrule
\begin{tabular}[c]{@{}c@{}}\textbf{AI}\\ \textbf{Integration}\end{tabular} & \begin{tabular}[c]{@{}c@{}}No native AI\\integration\end{tabular}
 & \begin{tabular}[c]{@{}c@{}}AI-assisted\\optimization (e.g., RAN)\end{tabular}
 & \begin{tabular}[c]{@{}c@{}}AI-driven network\\optimization and\\automation\end{tabular}
 & \begin{tabular}[c]{@{}c@{}}Fully AI-native,\\for autonomous,\\self-optimizing networks\end{tabular}
 \\ \midrule
\begin{tabular}[c]{@{}c@{}}\textbf{Network}\\ \textbf{Slicing}\end{tabular} & Not supported & \begin{tabular}[c]{@{}c@{}}Supported (e.g., cust-\\omized slices for IoT,\\ultra-low latency)\end{tabular}
 & \begin{tabular}[c]{@{}c@{}}Supported (e.g., cust-\\omized slices for IoT,\\ultra-low latency)\end{tabular}
 & \begin{tabular}[c]{@{}c@{}}Ultra-dynamic and\\adaptive slicing\\(real-time applications)\end{tabular} \\ \midrule
\begin{tabular}[c]{@{}c@{}}\textbf{Security}\\ \textbf{Mechanisms}\end{tabular} & \begin{tabular}[c]{@{}c@{}}Standard security\\(e.g., encryption)\end{tabular}
 & \begin{tabular}[c]{@{}c@{}}Enhanced security (e.g.,\\advanced encryption, 5G\\security architecture)\end{tabular}
 & \begin{tabular}[c]{@{}c@{}}Enhanced, dynamic\\security (e.g., \\ edge-based)\end{tabular}
 & \begin{tabular}[c]{@{}c@{}}Quantum encryption,\\AI-driven security\end{tabular}
 \\ \midrule
\begin{tabular}[c]{@{}c@{}}\textbf{Spectrum}\\\textbf{Efficiency}\end{tabular} & Moderate & High & \begin{tabular}[c]{@{}c@{}}NR, mmWave, 5G Core\\advanced aggregation)\end{tabular}& Extremely high, Terahertz utilization \\ \midrule
\begin{tabular}[c]{@{}c@{}}\textbf{Support}\\ \textbf{for}\\ \textbf{IoT}\end{tabular} & Limited & Massive IoT support & Optimized for massive IoT & \begin{tabular}[c]{@{}c@{}}Ubiquitous IoT support\\with ultra-low energy consumption\end{tabular}\\ \midrule
\begin{tabular}[c]{@{}c@{}}\textbf{Key}\\ \textbf{Technologies}\end{tabular} & OFDM, MIMO, LTE-A & \begin{tabular}[c]{@{}c@{}}NR, mmWave, 5G Core,\\Carrier Aggregation\end{tabular} & \begin{tabular}[c]{@{}c@{}}mmWave, Carrier Aggregation,\\Beamforming, Advanced MIMO \end{tabular} & \begin{tabular}[c]{@{}c@{}}AI/ML for network control, THz \\communication, Quantum computing\end{tabular} \\ \midrule
\begin{tabular}[c]{@{}c@{}}\textbf{Applications}\\ \textbf{and}\\ \textbf{Services}\end{tabular} & \begin{tabular}[c]{@{}c@{}}Mobile internet, HD\\video streaming, etc\end{tabular}
 & \begin{tabular}[c]{@{}c@{}}Connected cars, smart cities,\\telemedicine, remote work, etc \end{tabular} & \begin{tabular}[c]{@{}c@{}}8K video streaming, immersive,\\AR/VR, ultra-reliable\\low-latency communications, etc\end{tabular} & \begin{tabular}[c]{@{}c@{}}Hyper-connectivity, space\\communication, BCIs, etc\end{tabular} \\ 
\bottomrule
\end{tabular}%
}
\vspace{-0.2in}
\end{table*}
\begin{figure*}[htbp!]
\includegraphics[width=0.7\linewidth]{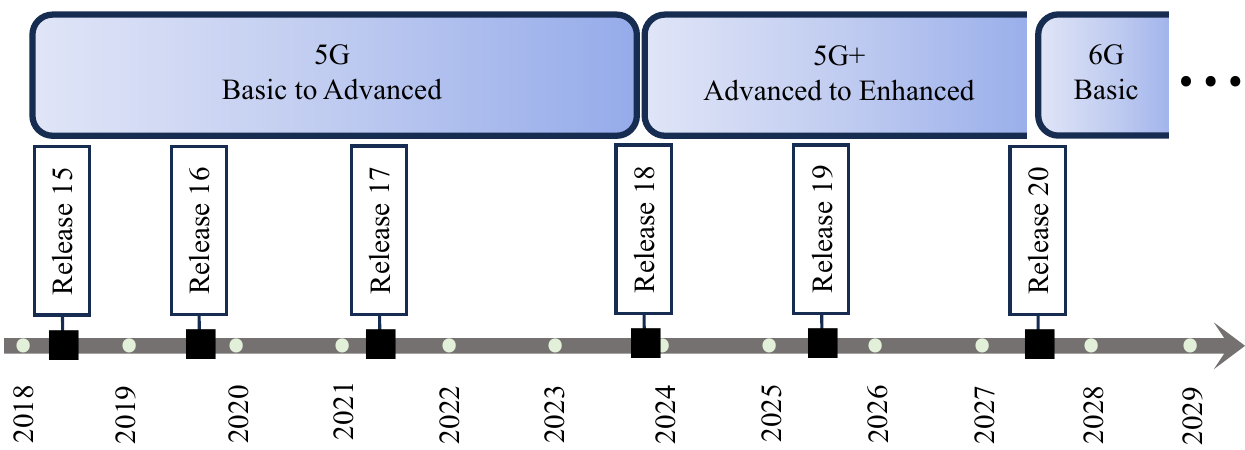}
\centering
\caption{From 3GPP to 5G/5G+ and 6G.}
\vspace{-0.2in}
\label{Fig9}
\end{figure*}
\subsection{Background and Literature Review}
\begin{figure*}[htbp!]
\includegraphics[width=0.7\linewidth]{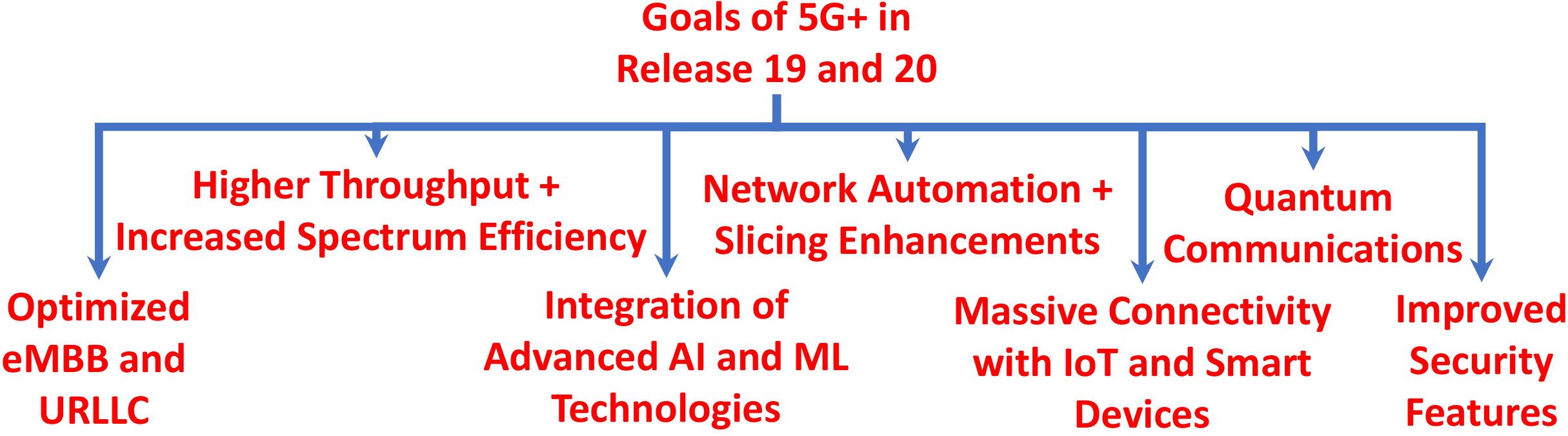}
\centering
\caption{Key goals of 5G+ in releases 18 and 19 in terms of efficiency, automation, AI integration, IoT, and security.}
\vspace{-0.2in}
\label{Fig10}
\end{figure*}

The integration of AI and ML into 5G and 5G+ wireless networks has become a crucial strategy for meeting the performance benchmarks outlined in 3GPP Releases 18 and 19 (see Fig.~\ref{Fig9} and \ref{Fig10}). Researchers have increasingly leveraged ML algorithms to enhance core network functions such as traffic classification, resource management \cite{KatoDeepLearningforNetworkTrafficControl2017, mao2016resource}, and power-efficient routing \cite{HuMLbasedAdaptiverouting2010}. RL and autoencoders have contributed to improved security \cite{rajathi2024adaptive, paper7JointTaskandData-OrientedSemantic2024} and traffic optimization \cite{taylor2023designing}, while unsupervised methods like K-means clustering have been applied to congestion control \cite{TaherMLbasedDatacongcont2016}. At the physical layer, DL continues to reshape traditional paradigms in channel estimation and symbol detection \cite{YeDLchannelestimation2018, demir2019channel}. DL-based approaches have demonstrated high performance with reduced complexity, offering robust signal reconstruction in impaired channels via autoencoders \cite{o2017introduction}. In terms of channel coding, Low-Density Parity-Check (LDPC) codes remain central to 5G implementations, particularly for eMBB scenarios \cite{fossorier2004quasicyclic}. However, further improvements in Bit Error Rate (BER) performance are needed, motivating AI-driven exploration in code design, decoding strategies, and adaptive error correction. \rc{Recent works \cite{paper9SwinJSCC2024, liu2023transformer} have also explored Transformer-based architectures for channel coding and decoding, using attention mechanisms to capture long-range dependencies and improve performance under complex channel conditions.}

While several surveys have explored AI's role across network layers \cite{zhang2020ai, morocho2019machine, akyildiz2020future}, recent work has emphasized Federated Learning (FL) for privacy-preserving distributed intelligence \cite{chen2020wireless, liu2020federated}, and RL and DL techniques for real-time optimization \cite{kaur2021machine, khan2020wireless}. Others have addressed channel coding in 5G systems \cite{arora2020survey, xu2020channel, ly2021review}, though often without deeply analyzing the intersection of AI and coding theory. Emerging perspectives highlight gaps in synthesizing AI-driven coding methods with evolving wireless infrastructure. For example, spectrum-sharing methods \cite{patil2024comprehensive} and UAV integration frameworks \cite{banafaa2024comprehensive} focus on deployment and efficiency but lack depth in exploring AI’s role in encoding strategies. As 5G progresses toward 6G and beyond, incorporating diverse spectral bands (mmWave, THz, optical) and advanced technologies like edge computing and quantum security, there is a critical need to align AI-enhanced coding mechanisms with the growing complexity of wireless systems.

Table~\ref{tab:survey_positioning_shortlist_restyled} compares this work with several recent, high-quality surveys/tutorials done recently. Existing surveys typically focus on (i) learning-based optimization at the network/system level, (ii) specific enabling technologies (e.g., cognitive radio, backscatter, blockchain), or (iii) upstream tasks such as channel modeling. In contrast, this survey is coding-centric and jointly addresses AI-driven channel coding/decoding and coding-aware resource optimization under explicit KPIs (reliability/BLER, latency, complexity, and energy).

\begin{table*}[!htbp]
\centering
\caption{Categorization and Feature Descriptions for MIMO Detection Algorithms}
\label{tab:mimo_detection}
\scriptsize
\renewcommand{\arraystretch}{1.2}
\setlength{\tabcolsep}{4pt}

\resizebox{\textwidth}{!}{%
\begin{tabular}{
>{\centering\arraybackslash}m{1.3cm}
>{\raggedright\arraybackslash}m{3.1cm}
>{\raggedright\arraybackslash}m{5.3cm}
>{\raggedright\arraybackslash}m{7.8cm}
}
\toprule
\textbf{Category} & \textbf{Type} & \textbf{Subcategory/Example} & \textbf{Description and Key Features} \\
\midrule

\rotatebox{90}{\textbf{Optimal}}
& \textbf{Exhaustive Search}
& \makecell[l]{-- Maximum Likelihood (ML) \\
-- Maximum a Posteriori (MAP)}
& \makecell[l]{-- Achieves optimal detection performance. \\
-- High complexity \& impractical for large-scale systems.} \\
\midrule

\multirow{13}{*}{\rotatebox{90}{\textbf{Suboptimal}}}
& \raisebox{-4.5mm}{\textbf{Linear Iterative}}
& \makecell[l]{-- Gauss--Seidel (GS) \\
-- Successive Over Relaxation (SOR) \\
-- Steepest Descent (SD) \\
-- Conjugate Gradient (CG) \\
-- Residual-Based (RBD) \\
-- Barzilai--Borwein (BB)}
& \makecell[l]{-- Iteratively refines the solution. \\
-- Offers a good complexity--performance trade-off.} \\
\cmidrule{2-4}

& \textbf{Linear}
& \makecell[l]{-- Matched Filter (MF) \\
-- Minimum Mean Square Error (MMSE) \\
-- Zero Forcing (ZF)}
& \makecell[l]{-- Simple non-iterative techniques. \\
-- Low complexity but limited performance in large systems.} \\
\cmidrule{2-4}

& \textbf{Nonlinear}
& \makecell[l]{-- Interference Cancellation (IC) \\
-- Successive IC (SIC), Parallel IC (PIC) \\
-- Tree Search: DFS, BFS \\
-- Message Passing: AMP, OAMP, BP, EP}
& \makecell[l]{-- Employs nonlinear processing for improved robustness. \\
-- Outperforms linear methods with variable complexity.} \\
\cmidrule{2-4}

& \textbf{AI-Based}
& \makecell[l]{-- ComNet \\
-- DetNet \\
-- MMNet \\
-- DNN-based Message Passing}
& \makecell[l]{-- Utilizes machine/DL techniques. \\
-- Learns near-optimal detection strategies. \\
-- High performance but requires training and adaptation.} \\
\bottomrule
\end{tabular}%
}
\end{table*}

\begin{table*}[htbp!]
\centering
\caption{Positioning of this survey in comparison with the recent surveys on AI/ML for wireless communications}
\label{tab:survey_positioning_shortlist_restyled}
\renewcommand{\arraystretch}{1}
\setlength{\tabcolsep}{4pt}
\scriptsize

\begin{tabularx}{\textwidth}{
p{1cm}
p{0.8cm}
p{2.1cm}
p{3.0cm}
X
}
\hline
\textbf{Ref (venue)} &
\textbf{Year} &
\textbf{Primary Scope} &
\textbf{Main Coverage} &
\textbf{How \emph{this} survey differs (Unique value-add)} \\
\hline

\cite{ShiCOMST2023_L2O_6G} IEEE COMST) &
2023 &
ML for large-scale wireless optimization (learning to optimize) &
DL/RL-based optimization frameworks for 6G; scalability, convergence, and system-level objectives &
Extent beyond optimization-only perspectives by introducing a \emph{coding-aware reliability layer}: AI-driven FEC decoding and code design, and how coding decisions reshape optimization tradeoffs (e.g., redundancy vs.\ retransmissions, compute vs.\ latency), which are outside the scope of L2O-centric surveys. \\

\cite{MahmoodACCESS2022_IoT6G_AIReview} (IEEE Access) &
2022 &
AI/ML for IoT–6G &
High-level survey of ML methods and use cases in IoT/6G &
Shifts from broad surveys to a \emph{focused, technical treatment} of learned coding/decoding, emphasizing reliability KPIs (BLER, latency, complexity, energy) and practical decoder deployment. \\

\cite{AIEmpoweredBackscatter_IoTJ_2023} (IEEE IoT-J) &
2023 &
AI-enabled backscatter communications &
DL/RL methods for backscatter system design, energy efficiency, and reliability &
Extends beyond backscatter by covering mainstream error-control codes and their interaction with resource allocation decisions, and by identifying coding-centric open challenges that are missing in \emph{backscatter device-focused} surveys. \\

\cite{JSAC2024_OptMethodsWirelessSurvey} (IEEE JSAC) &
2024 &
Optimization methods for wireless communications &
Algorithmic and optimization-theoretic methods; AI as a supporting tool &
Adds a \emph{learning-for-coding} dimension: neural and unfolded decoders, AI-assisted code construction, and reliability-driven adaptation with explicit BLER--latency--energy coupling, which are not treated as first-class topics in optimization-method surveys. \\

\cite{LyYao_OJCOMS_2021_DL5GReview} (IEEE OJ-COMS) &
2021 &
Broad DL survey for 5G research &
DL applications across LDPC channel coding, massive MIMO, NOMA, resource allocation, and network security &
Moves beyond generation-specific surveys to a \emph{coding-centric and reliability-driven} view that integrates AI-based coding/decoding with KPI-aware resource optimization, emphasizing deployability and system-level optimization.\\

\cite{BlockchainAI6G_COMST2023} (IEEE COMST) &
2023 &
Blockchain + AI for 6G &
AI--blockchain integration for security, trust, and service enablement &
Keeps security and trust as supporting context, but centers \emph{reliability via coding/decoding}, offering a unified coding-aware optimization narrative (latency/compute/energy under reliability targets) absent in blockchain-centric surveys. \\

\cite{AIEnergyOpt_RAN_Survey} (IEEE Access) &
2024 &
AI-driven energy optimization in RAN &
Learning-based energy saving and OPEX/emission reduction in RAN &
Extends energy-only optimization to \emph{multi-KPI tradeoffs} by incorporating reliability (BLER) and decoder complexity/latency, and by introducing coding-aware levers such as adaptive redundancy and learned decoding complexity control. \\

\cite{MLCognitiveRadio_COMST2024} (IEEE COMST) &
2024 &
ML-driven cognitive radio &
Learning-based spectrum sensing, access, and adaptation &
Complements spectrum-centric CR surveys with a \emph{reliability-first} view, emphasizing coding/decoding and their interaction with spectrum and resource constraints.\\

\cite{Waqas_WirelessSecurity_AI_Review} (AI Review) &
2022 &
AI/ML for wireless security &
Survey of ML/DL/RL methods for authentication, intrusion detection, jamming, malware, and traffic analysis &
Restricted to security mechanisms and \emph{excludes} coding design and decoding, our survey instead treats error-control coding as a core component and analyzes its impact on learning-based resource optimization trade-offs. \\

\cite{Alhammadi_AI_6G_IJIS_2024} (IJISs) &
2025 &
Broad AI for 6G wireless networks &
Broad overview of AI methods applied to diverse 6G use cases, including resource management, MIMO, RIS, channel modeling, slicing, security, and emerging technologies. &
Gives a broad AI–6G overview where coding is only briefly mentioned. In contrast, our survey is coding-centric, analyzing AI-based code design and decoding, and their impact on learning-driven resource optimization and system KPIs.\\

\cite{AIProp_ChannelModel_TAP_2022} (IEEE TAP) &
2022 &
AI-enabled propagation and channel modeling &
ML-based scenario identification and channel characterization &
Positions channel modeling as an upstream enabler, while contributing to the downstream reliability layer: modern coding/decoding advances (LDPC, polar, neural BP/unfolding) and coding-aware optimization under realistic deployment constraints. \\

\textbf{This work} &
-- &
AI-driven \emph{channel coding/decoding} with \emph{coding-aware resource optimization} &
Learned decoding and code construction (DL/RL, model-driven and data-driven); adaptive redundancy; KPI-aware optimization across latency, reliability (BLER), energy, and compute; deployability considerations &
Provides a unified, \emph{coding-centric} synthesis that jointly addresses (i) learned decoding, code construction, and adaptive redundancy, (ii) explicit coupling between coding reliability and system-level resource optimization (latency--energy--compute tradeoffs), and (iii) deployment constraints (generalization, complexity, robustness), without anchoring novelty to a single wireless generation. \\
\hline

\end{tabularx}
\end{table*}
\rc{
\subsection{Survey Scope and Literature Selection Methodology}
To ensure transparency, this survey adopts a structured literature selection process. Relevant studies were collected from IEEE Xplore, Scopus, Web of Science, ScienceDirect, and Google Scholar using keywords such as AI, ML, DL, RL, channel coding, LDPC, polar codes, resource optimization, MIMO, NOMA, 5G, and 5G+. The review mainly covers studies published from 2020 to 2025, with a few earlier foundational works included where necessary. Studies were included if they: (i) were directly related to AI-driven channel coding, coding-aware resource optimization, or related physical-layer wireless communication tasks; (ii) appeared in peer-reviewed journals, conferences, or high-quality surveys/tutorials; and (iii) provided enough technical detail for comparison. Duplicates, marginally relevant papers, non-technical opinion articles, and studies without a clear methodological contribution were excluded. The overall selection process is illustrated in Fig. \ref{survey_selection_workflow}. Although this work is not intended as a statistical meta-analysis, it follows a structured and transparent survey methodology to improve reproducibility and scope definition.}
\begin{figure*}[t]
\centering
\resizebox{0.88\textwidth}{!}{%
\begin{tikzpicture}[
    font=\small,
    >=Latex,
    node distance=0.9cm and 0.9cm,
    mainbox/.style={
        rectangle,
        rounded corners=3pt,
        draw=#1!70!black,
        very thick,
        fill=#1!12,
        text width=4.25cm,
        minimum height=2.35cm,
        align=center,
        inner sep=6pt
    },
    exbox/.style={
        rectangle,
        rounded corners=3pt,
        draw=#1!70!black,
        thick,
        dashed,
        fill=#1!10,
        text width=4.95cm,
        minimum height=2.75cm,
        align=left,
        inner sep=6pt
    },
    flow/.style={->, very thick, draw=black!75},
    sideflow/.style={->, thick, draw=black!65}
]

\colorlet{idcol}{blue}
\colorlet{scrcol}{teal}
\colorlet{elicol}{orange}
\colorlet{inccol}{green!55!black}
\colorlet{excol}{red!65!black}

\node[mainbox=idcol] (id) {
\textbf{\large Identification}\\[2pt]
Records identified through database search\\
\textit{(IEEE Xplore, Scopus, Web of Science, ScienceDirect, Google Scholar)}\\
$Articles=\textit{374}$
};

\node[mainbox=scrcol, right=1.0cm of id] (scr) {
\textbf{\large Screening}\\[2pt]
Records after duplicate etc. removal\\
$Articles=\textit{374}$
};

\node[mainbox=elicol, right=1.0cm of scr] (elig) {
\textbf{\large Eligibility}\\[2pt]
Full-text articles assessed for eligibility\\
$Articles=\textit{253}$
};

\node[mainbox=inccol, right=1.0cm of elig] (incl) {
\textbf{\large Included Studies}\\[2pt]
Studies included in the final survey synthesis\\
$Articles=\textit{169}$\\[2pt]
\textit{Primary focus: [2020--2025]; seminal earlier studies retained where needed}
};

\node[exbox=excol, below=1.35cm of scr] (ex1) {
\textbf{Excluded after title/abstract screening}\\
$Articles=\textit{121}$\\
-- Out of scope\\
-- Duplicate/overlapping studies\\
-- Non-peer-reviewed articles\\
-- Irrelevant to AI-driven channel coding or resource optimization
};

\node[exbox=excol, below=1.35cm of elig] (ex2) {
\textbf{Excluded after full-text review}\\
$Articles=\textit{84}$\\
-- Insufficient technical depth\\
-- Peripheral relevance to survey scope\\
-- Unclear methodological contribution\\
-- Not aligned with 5G/5G+ wireless focus
};

\draw[flow] (id.east) -- (scr.west);
\draw[flow] (scr.east) -- (elig.west);
\draw[flow] (elig.east) -- (incl.west);

\draw[sideflow] (scr.south) -- (ex1.north);
\draw[sideflow] (elig.south) -- (ex2.north);

\node[font=\small, text=black!60, right=3pt]
    at ($(scr.south)!0.5!(ex1.north)$) {Screened out};

\node[font=\small, text=black!60, right=3pt]
    at ($(elig.south)!0.5!(ex2.north)$) {Excluded};

\end{tikzpicture}%
}
\caption{\rc{Literature selection workflow adopted in this survey, showing identification, screening, eligibility assessment, and final inclusion of relevant studies.}}
\label{survey_selection_workflow}
\end{figure*}

\section{AI-Driven Detection and Precoding Techniques for MIMO Systems}\label{AI-DrivenDetection}
\rc{AI-based MIMO detection and precoding methods are most beneficial when channel models are inaccurate, interference is strong, or iterative optimization becomes too costly; however, their gains depend on training distribution, CSI quality, and inference complexity}. DL models enable accurate and efficient signal detection in complex environments, reducing computational demands. For precoding, AI-driven techniques optimize resource allocation, energy use, QoS, and beamforming, enhancing spectral efficiency and communication reliability.

\subsection{AI Solutions for Detection in MIMO Systems}

\subsubsection{Overview}
To improve the efficiency of MIMO detection systems, a model-driven DL strategy merges iterative signal processing algorithms with DNNs \cite{ModelDriven2020}. In this framework, each DNN layer emulates an iteration of the algorithm, with trainable parameters integrated to refine detection accuracy. Consider a narrowband MIMO system with \( M_t \) transmit and \( M_r \) receive antennas, modeled as,

\begin{equation}\label{eq1}
\mathbf{r} = \mathbf{G} \mathbf{x} + \mathbf{z},
\end{equation}
where \( \mathbf{r} \in \mathbb{C}^{M_r} \) denotes the received signal vector, \( \mathbf{G} \in \mathbb{C}^{M_r \times M_t} \) is the channel matrix, \( \mathbf{x} \in \Phi^{M_t} \) represents the transmitted symbols drawn from constellation \( \Phi \), and \( \mathbf{z} \sim \mathcal{CN}(0, \eta^2 \mathbf{I}_{M_r}) \) is additive white Gaussian noise (AWGN). While maximum likelihood detection achieves optimal performance, its computational cost grows exponentially with antenna count or modulation order, rendering it impractical for large-scale systems. Consequently, approximate detection algorithms categorized as optimal and suboptimal, described in Table~\ref{tab:mimo_detection} are employed. 

Linear detection methods, such as ZF \cite{zeroforcing2015} and Minimum Mean Squared Error (MMSE) \cite{MMSEequalization2023}, are widely used due to their simplicity and low computational complexity. The ZF estimator suppresses interference by inverting the channel matrix,

\begin{equation}
    \hat{\mathbf{x}}_{\text{ZF}} = (\mathbf{G}^H \mathbf{G})^{-1} \mathbf{G}^H \mathbf{r},
\end{equation}
where \( (\cdot)^H \) denotes the Hermitian transpose. However, the performance of ZF deteriorates in noisy environments, particularly when the channel matrix \( \mathbf{G} \) is ill-conditioned, as the inversion process amplifies noise. 

To address this limitation, MMSE detection incorporates the noise variance \( \eta^2 \) into the inversion process. The MMSE estimate is given by,

\begin{equation}
    \hat{\mathbf{x}}_{\text{MMSE}} = (\mathbf{G}^H \mathbf{G} + \eta^2 \mathbf{I})^{-1} \mathbf{G}^H \mathbf{r},
\end{equation}
where \( \mathbf{I} \) is the identity matrix and \( \eta^2 \) represents the noise variance. The regularization term \( \eta^2 \mathbf{I} \) stabilizes the inversion, enhancing robustness in low Signal-to-Noise Ratio (SNR) scenarios. This approach balances interference suppression and noise amplification, minimizing the overall Mean Squared Error (MSE), defined as,

\begin{equation}
    \text{MSE} = \mathbb{E}[\|\mathbf{x} - \hat{\mathbf{x}}\|^2].
\end{equation}

Despite its advantages, MMSE detection also faces challenges under complex propagation conditions, such as multipath fading or highly correlated channels, where its performance may degrade.

Linear iterative methods, including Gauss-Seidel (GS) \cite{ahmadi2021parallel}, Coordinate Descent (CD) \cite{nie2021coordinate}, Successive Over Relaxation (SOR) \cite{wadayama2021chebyshev}, and Steepest Descent (SD) \cite{cocchi2020convergence}, refine linear detection techniques by iteratively updating the solution. These methods address the limitations of direct approaches like ZF and MMSE, offering improved performance in certain scenarios.

For instance, GS solves the linear system \( \mathbf{G} \mathbf{x} = \mathbf{r} \) by sequentially updating each component of \( \mathbf{x} \). The update for the \( i \)-th component at iteration \( k+1 \) is denoted by $x_i$ and expressed by,

\begin{equation}
x_i^{(k+1)} = \frac{1}{\mathbf{G}_{ii}} \left( r_i - \sum_{j < i} \mathbf{G}_{ij} x_j^{(k+1)} - \sum_{j > i} \mathbf{G}_{ij} x_j^{(k)} \right).
\end{equation}

Similarly, CD minimizes a quadratic cost function by optimizing one coordinate at a time. For a cost function defined as,

\begin{equation}
f(\mathbf{x}) = \frac{1}{2} \mathbf{x}^T \mathbf{Q} \mathbf{x} - \mathbf{b}^T \mathbf{x},
\end{equation}
and the update for the \( i \)-th coordinate is expressed as,

\begin{equation}
x_i^{(k+1)} = x_i^{(k)} - \frac{1}{\mathbf{Q}_{ii}} \frac{\partial f(\mathbf{x})}{\partial x_i} \bigg|_{\mathbf{x} = \mathbf{x}^{(k)}}.
\end{equation}

SOR enhances GS by introducing a relaxation factor \( \omega \) to accelerate convergence. The update rule for SOR is defined as,
\begin{equation}
x_i^{(k+1)} = (1 - \omega) x_i^{(k)} + \frac{\omega}{\mathbf{G}_{ii}} \Biggl( r_i - \sum_{j < i} \mathbf{G}_{ij} x_j^{(k+1)}-\sum_{j > i} \mathbf{G}_{ij} x_j^{(k)} \Biggr).
\end{equation}

Finally, SD iteratively minimizes the residual \( \mathbf{r}^{(k)} = \mathbf{r} - \mathbf{G} \mathbf{x}^{(k)} \) by updating \( \mathbf{x} \) along the steepest gradient direction. The update rule for SD is,

\begin{equation}
    \mathbf{x}^{(k+1)} = \mathbf{x}^{(k)} + \alpha^{(k)} \mathbf{r}^{(k)},
\end{equation}
where the step size \( \alpha^{(k)} \) is computed as,

\[
\alpha^{(k)} = \frac{\|\mathbf{r}^{(k)}\|^2}{\|\mathbf{G} \mathbf{r}^{(k)}\|^2}.
\]

While these iterative methods improve upon the simplicity of ZF and MMSE, their performance often remains suboptimal in realistic propagation conditions, such as highly correlated or time-varying channels. Enhancing their convergence and robustness under dynamic channel conditions remains an active area of research.

To overcome detection challenges in MIMO systems, nonlinear algorithms such as Sphere Decoding (SD) have been developed for optimal performance at the cost of high complexity. SD uses a tree search to solve,
\begin{equation}
\hat{\mathbf{x}} = \arg \min_{\mathbf{x} \in \Phi^{M_t}} \|\mathbf{r} - \mathbf{G} \mathbf{x}\|^2,
\end{equation}
where \( \Phi \) is the modulation alphabet, \( \mathbf{G} \) the channel matrix, \( \mathbf{r} \) the received signal, and \( \mathbf{x} \) the transmitted signal vector \cite{nguyen2021application}. While effective, SD's complexity scales exponentially with the number of antennas (\( M_t \)) and modulation order, making it unsuitable for large systems.

Interference Cancellation (IC) techniques, such as Successive Interference Cancellation (SIC), iteratively subtract detected signals to reduce interference,

\begin{equation}
\mathbf{r}^{(k+1)} = \mathbf{r}^{(k)} - \mathbf{G} \hat{\mathbf{x}}^{(k)},
\end{equation}
where \( \hat{\mathbf{x}}^{(k)} \) is the detected symbol in iteration \( k \) \cite{kong2020frame}. These methods depend heavily on accurate channel knowledge and correct detection order.

Message Passing (MP) algorithms use factor graphs to estimate marginal probabilities \( P(x_i | \mathbf{r}) \) through iterative message exchange \cite{feng2022unifying}. Though theoretically strong, MP often underperforms in correlated or fast-fading environments.

Despite their performance benefits, nonlinear methods remain constrained by complexity and sensitivity to channel variations. To address these issues, AI-based MIMO detection offers a compelling alternative. DL models capture complex, nonlinear signal relationships, adapt to changing channel conditions, and reduce reliance on precise mathematical models. These solutions also scale better in large systems, bridging the gap left by traditional techniques. The comparative strengths and weaknesses of MIMO detection methods are summarized in Table~\ref{tab:mimo_detectors2}.

\subsubsection{Linear Detection Based on DL}
DL has emerged as a powerful enhancement to linear detection in MIMO systems, addressing limitations of traditional methods like ZF and MMSE, which degrade under correlated signals or harsh propagation environments.

Detection Network (DetNet) \cite{LearningtoDetect2019} improves detection by framing it as an iterative learning task. Given the model in \eqref{eq1}, where \( \mathbf{r} \) is the received signal, \( \mathbf{G} \) the channel matrix, and \( \mathbf{x} \) the transmitted vector, DetNet refines the estimate \( \hat{\mathbf{x}}_k \) at each layer using projected gradient descent,

\begin{equation}
\begin{aligned}
\mathbf{p}_k &= \hat{\mathbf{x}}_{k-1} - \phi_k^{(1)} \mathbf{G}^H \mathbf{r} + \phi_k^{(2)} \mathbf{G}^H \mathbf{G} \mathbf{x}_{k-1}, \\
\mathbf{q}_k &= \text{ReLU}\left(\Phi_k^{(1)}\left[\begin{array}{c}
\mathbf{p}_k \\ \mathbf{w}_{k-1}
\end{array}\right] + \phi_k^{(1)}\right), \\
\hat{\mathbf{x}}_k &= \Phi_k^{(2)} \mathbf{q}_k + \phi_k^{(2)}, \quad
\mathbf{w}_k = \Phi_k^{(3)} \mathbf{q}_k + \phi_k^{(3)},
\end{aligned}
\end{equation}
where ReLU is the rectified linear unit and \( \{\phi_k^{(1)}, \phi_k^{(2)}, \Phi_k^{(1)}, \Phi_k^{(2)}, \Phi_k^{(3)}\} \) are trainable parameters. While DetNet performs well under ideal (i.i.d. Gaussian) channels, its high complexity and limited adaptability to realistic MIMO models reduce practicality.

\begin{table*}[htbp!]
\centering
\caption{Comparison of MIMO Detectors}
\label{tab:mimo_detectors2}
\scriptsize
\begin{tabular}{ll}
\toprule
\textbf{Technique}       & \textbf{Advantages / Limitations} \\ \midrule
\textbf{Optimal}         & \begin{tabular}[c]{@{}l@{}}Advantages: Optimal detection performance \\ Limitations: NP-hard, impractical for large systems\end{tabular} \\ \midrule
\textbf{Tree-Search Based} & \begin{tabular}[c]{@{}l@{}}Advantages: Near-optimal, reduced complexity \\ Limitations: Exponential complexity for massive systems\end{tabular} \\ \midrule
\textbf{Linear}          & \begin{tabular}[c]{@{}l@{}}Advantages: Low complexity, good for small MIMO \\ Limitations: Requires \(O(N^3)\) matrix inversion; limited for large MIMO\end{tabular} \\ \midrule
\textbf{Linear Iterative} & \begin{tabular}[c]{@{}l@{}}Advantages: \(O(N^2)\) complexity; simple implementation \\ Limitations: Performance limited by MMSE\end{tabular} \\ \midrule
\textbf{Message Passing} & \begin{tabular}[c]{@{}l@{}}Advantages: Soft output, good performance, reconfigurable \\ Limitations: Convergence issues; hard-to-optimize factors\end{tabular} \\ \midrule
\textbf{AI Based}        & \begin{tabular}[c]{@{}l@{}}Advantages: Learns and adapts, improved performance \\ Limitations: High training complexity; re-training needed\end{tabular} \\ \bottomrule
\end{tabular}%
\end{table*}

To improve generalization, ComNet \cite{ComNet2018} combines ZF-based preprocessing with neural refinement. This hybrid design merges classical efficiency with data-driven accuracy, achieving modest yet interpretable gains.

Similarly, in \cite{DLBasedMIMOOFDMDetector2022}, a deep model approximates the output of the Conjugate Gradient-enhanced Orthogonal Approximate Message Passing (OAMP) algorithm, reducing the cost of matrix inversions. The network minimizes,

\begin{equation}
L(\theta) = \frac{1}{N} \sum_{i=1}^{N} \left\| r_i - \mathbf{w}^T x_i \right\|^2,
\end{equation}
where \( r_i \) is the received signal, \( x_i \) the input, and \( \mathbf{w} \) the learned weights.

Similarly, \cite{DLMassiveMIMOUplink2022} employs DL to directly estimate transmitted symbols in massive MIMO. The model is trained with,

\begin{equation}
L(\mathbf{W}, \mathbf{r}, \mathbf{X}) = \frac{1}{N} \sum_{i=1}^{N} \left\| \mathcal{N}(\mathbf{r}_i) - \mathbf{X}_i \right\|^2,
\end{equation}
where \( \mathcal{N}(\cdot) \) is the network output and \( \mathbf{X}_i \) the ground truth.

These studies confirm that DL enhances detection performance and scalability in MIMO systems while reducing reliance on complex linear algebra.

\subsubsection{Message Passing Algorithms Inspired by DL}
Message-passing algorithms are essential for MIMO detection, with Approximate Message Passing (AMP) offering simplicity and efficiency. However, AMP’s reliance on i.i.d. Gaussian channel matrices limits real-world applicability. To address this, Orthogonal AMP (OAMP) extends AMP to non-Gaussian settings \cite{mimoapm2019}.

OAMPNet \cite{ModelDriven2020} builds on OAMP by introducing trainable parameters \( \{\gamma_k, \theta_k\} \) for dynamic tuning at each iteration,

\begin{equation}
\begin{aligned}
    \mathbf{t}_k &= \hat{\mathbf{x}}_k + \gamma_k \mathbf{V}_k ( \mathbf{r}' - \mathbf{G} \hat{\mathbf{x}}_k ), \\
    \hat{\mathbf{x}}_{k+1} &= \mathbb{E}\{ \mathbf{x}' \mid \mathbf{t}_k, \lambda_k^2(\theta_k) \},
\end{aligned}
\end{equation}
where \( \mathbf{V}_k \) is the linear MMSE matrix. While OAMPNet improves accuracy, it assumes unitary-invariant channels and requires matrix inversions, limiting scalability.

To overcome these, MMNet \cite{adaptiveNNdectection2020} introduces trainable matrices and scalars without assuming unitary invariance,
\begin{equation}
\begin{aligned}
    \mathbf{t}_k &= \hat{\mathbf{x}}_k + \Psi_k^{(1)} ( \mathbf{r}' - \mathbf{G} \hat{\mathbf{x}}_k ), \\
    \hat{\mathbf{x}}_{k+1} &= \mathbb{E}\{ \mathbf{x}' \mid \mathbf{t}_k, \lambda_k^2(\psi_k^{(2)}) \},
\end{aligned}
\end{equation}
where \( \Psi_k^{(1)} \) and \( \psi_k^{(2)} \) are learnable. MMNet eliminates matrix inversion and supports online retraining for dynamic environments, achieving better performance and lower complexity than OAMPNet.

Belief Propagation (BP) algorithms also benefit from DL. Variants like damped BP (DNN-dBP) and max-sum BP (DNN-MS) \cite{MassiveMIMOMessage2020} integrate neural networks to refine message passing, improving robustness under imperfect CSI.

\begin{figure*}[htbp!]
\includegraphics[width=0.6\linewidth]{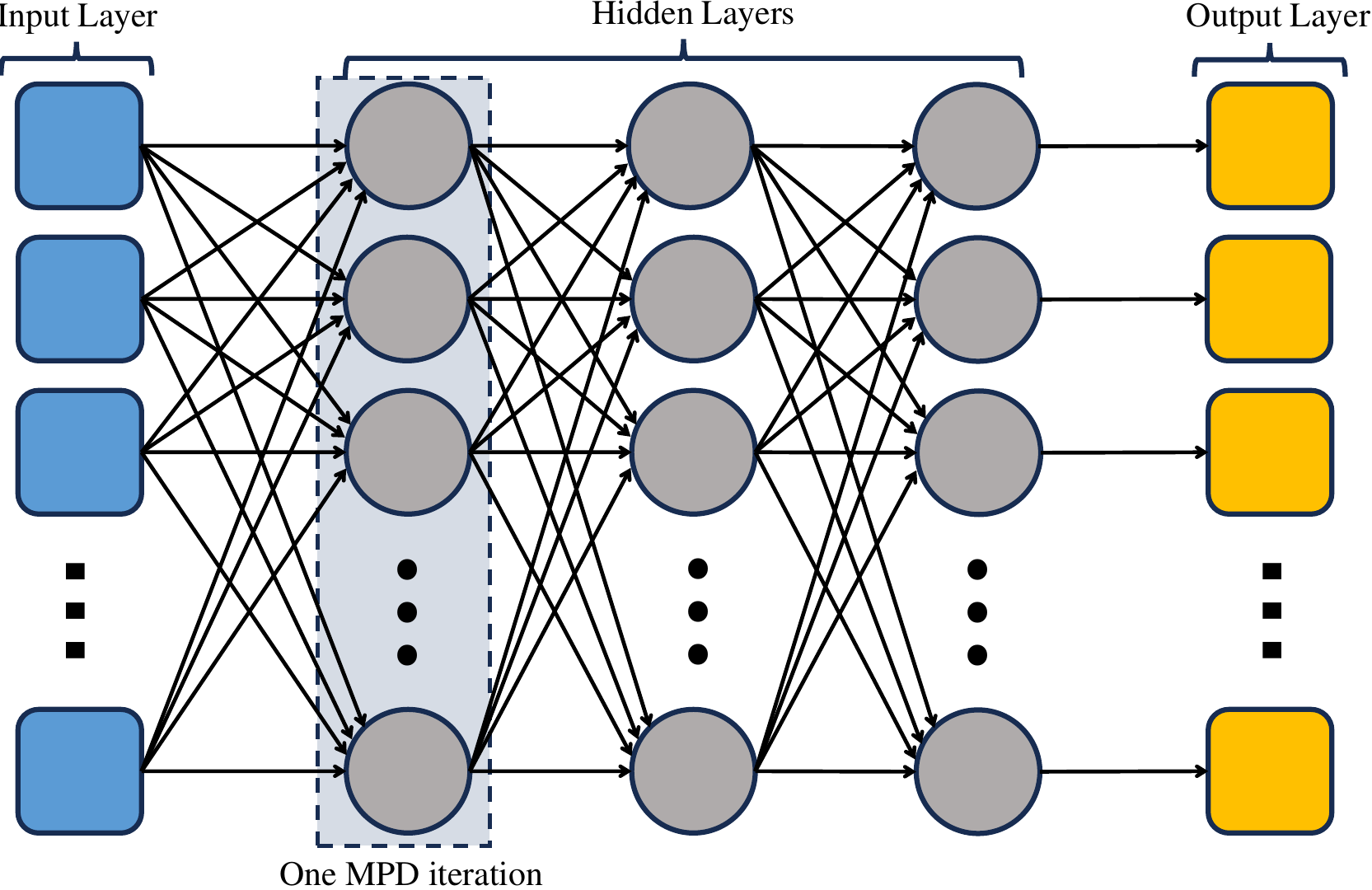}
\centering
\caption{Generic DNN-based MPD architecture showing key design principles, \rc{adapted from \cite{MassiveMIMOMessage2020}.}}
\label{Fig2}
\end{figure*}

The DNN-based Message-Passing Detector (MPD) architecture is shown in Fig.~\ref{Fig2}. It includes: (1) an input layer (blue squares) for noisy signals, (2) fully connected hidden layers (gray circles) simulating MPD iterations, and (3) an output layer (yellow squares) for symbol estimates. \rc{While generic, this architecture captures the key design components shared across recent learning-based MPD approaches reviewed in this paper. Its scalable design allows it to adapt to a wide range of MIMO system sizes.}

Trained on real data, these architectures replicate and improve iterative detection, achieving superior performance and scalability in large MIMO systems.

\subsection{AI in Multi-User MIMO Precoding}
Multi-user (MU) MIMO systems, especially in large-scale deployments, face significant challenges in designing precoding strategies that effectively manage interference while adhering to power constraints. Precoding, a critical operation in the downlink of MU-MIMO, determines how signals are spatially processed and transmitted to multiple user equipment. In a narrowband MU-MIMO system, the interaction between the received and transmitted signals is expressed similarly to model \eqref{eq1}. The goal of precoding is to optimize \( \mathbf{x} \) so that each User Equipment (UE) receives its desired signal with minimal interference and noise. Linear precoding methods, such as the Wiener filter precoder \( \mathbf{Q}_{\text{WF}} \), aim to minimize the MSE at the UEs. The Wiener filter is given by,

\begin{equation}
    \mathbf{Q}_{\text{WF}} = \left( \mathbf{G}^H \mathbf{G} + \frac{M N_0}{\rho^2} \right)^{-1} \mathbf{G}^H,
\end{equation}
where \( \mathbf{G}^H \) is the Hermitian transpose of the channel matrix, \( \rho^2 \) is the transmit power constraint, \( M \) represents the number of UEs, and \( N_0 \) is the noise variance. Although linear precoding is computationally efficient, it often fails to adequately manage interference and achieve optimal performance, particularly in high-density deployments or under complex propagation conditions.

To overcome these limitations, nonlinear precoding methods have been developed. These methods typically involve iterative optimization techniques that can be computationally intensive. AI has emerged as a transformative tool to enhance nonlinear precoding by taking advantage of data-driven methods to learn and optimize the required parameters. One prominent approach integrates AI with iterative optimization, refining the precoded signal over multiple iterations \cite{1bitMassiveMUMIMO2017}. The process can be expressed as,

\begin{equation}
\begin{aligned}
    \mathbf{v}^{(t+1)} &= \mathbf{x}^{(t)} - \tau^{(t)} \mathbf{F}^H \mathbf{F} \mathbf{x}^{(t)}, \\
    \mathbf{x}^{(t+1)} &= \operatorname{prox}\left( \mathbf{v}^{(t+1)}; \eta^{(t)}, \xi^{(t)} \right),
\end{aligned}
\end{equation}
where \( \mathbf{F} \) is a modified version of the channel matrix, \( \tau^{(t)} \) is a learnable step size parameter, and the proximal operator \( \operatorname{prox} \) adjusts \( \mathbf{v}^{(t+1)} \) by clipping the real and imaginary parts of the signal to a specified range \( [-\xi, +\xi] \). The parameters \( \eta^{(t)} \) and \( \xi^{(t)} \) control the range of this clipping operation. By optimizing these parameters using ML, the system can adaptively adjust the precoding process to varying channel conditions and UE requirements.

Deep unfolding techniques offer a promising framework for combining iterative optimization with AI. These methods treat each iteration of the optimization as a layer in a neural network, allowing the model to learn optimal parameter settings \( \{ \tau^{(t)}, \eta^{(t)}, \xi^{(t)} \} \) during training. This approach reduces the number of required iterations while maintaining or even improving performance. Furthermore, deep unfolding enables real-time adaptation to dynamic channel states, making it particularly well-suited for practical deployments where channel conditions can change rapidly.

\section{AI in Coding Theory for 5G/5G+ Networks}\label{AIinCodingTheory}
\subsection{Brief Overview of Coding Theory}
In wireless communications, error correction is vital due to interference, fading, and noise. Coding theory enhances reliability and channel capacity, the maximum achievable error-free transmission rate. For 5G/5G+ networks, which demand high throughput, low latency, and efficient power usage, advanced coding schemes must balance redundancy with data rate efficiency. Key coding techniques include algebraic block codes, convolutional codes, turbo codes, LPDC codes, fountain codes, and polar codes. Among these, LDPC and polar codes are selected for 5G due to their superior error correction performance \cite{arora2020survey}. Algebraic block codes, while effective, require full message reception for decoding. Convolutional codes, used in 3G systems, improve BER but at higher complexity \cite{zheng2008beyond}. Turbo codes approach Shannon’s limit, maximizing transmission efficiency \cite{costello2007channel}. Rateless codes like LT codes \cite{luby2002lt} and fountain codes \cite{mehran2015rateless} excel in erasure channels, making them ideal for storage and multicast scenarios.

Polar codes, introduced by Arikan, offer low-complexity encoding/decoding with near-capacity performance, making them suitable for 5G applications \cite{bioglio2020design}. LDPC codes, defined by sparse parity-check matrices, also provide strong error correction. Quasi-cyclic LDPC variants are memory-efficient and scalable, aligning with 5G’s need for flexibility \cite{ranganathan2019quasi}.

\subsection{Decoding in Polar Codes and LDPC Codes}
LDPC and Polar codes are central to modern wireless communication systems, offering the error correction and efficiency required by 5G and beyond \cite{jang2021improving}.

\subsubsection{LDPC Codes}
LDPC codes are characterized by sparse parity-check matrices, which enable efficient error correction near the Shannon limit. The decoding process generally employs iterative techniques like the BP algorithm, which facilitates message passing between variable nodes and check nodes in a Tanner graph. The message update in BP can be expressed as,

\begin{equation}
    L_{v \to c} = L_{ch} + \sum_{c' \in N(v) \setminus c} L_{c' \to v},
\end{equation}
where \( L_{v \to c} \) represents the message from a variable node \( v \) to a check node \( c \), \( L_{ch} \) represents the channel likelihood, and \( N(v) \) is the set of check nodes neighboring variable node \( v \). Similarly, the check-to-variable node update is given by,

\begin{equation}
L_{c \to v} = 2 \tanh^{-1} \left( \prod_{v' \in N(c) \setminus v} \tanh\left(\frac{L_{v' \to c}}{2}\right) \right),
\end{equation}
where \( L_{c \to v} \) refers to the message transmitted by the check node \( c \) to variable node \( v \). While the BP algorithm is effective, it is computationally demanding. To mitigate this, the min-sum approximates the BP updates by replacing the hyperbolic tangent with a minimum operation,

\begin{equation}
    L_{c \to v} \approx \min_{v' \in N(c) \setminus v} \left| L_{v' \to c} \right| \cdot \prod_{v' \in N(c) \setminus v} \operatorname{sgn}(L_{v' \to c}).
\end{equation}

Layered decoding further optimizes performance by processing subsets of the Tanner graph sequentially, significantly reducing latency and improving convergence, which is particularly important for real-time applications in 5G systems.

\subsubsection{Polar Codes}
Polar codes are specifically constructed to reach the capacity of symmetric binary-input discrete memoryless channels through a technique referred to as channel polarization. The structure of polar codes is defined recursively, with the generator matrix \( G_N \) given by,

\begin{equation}
G_N = B_N F^{\otimes n},
\end{equation}
where \( N = 2^n \), \( B_N \) is the bit-reversal permutation matrix, and \( F = \begin{bmatrix} 1 & 0 \\ 1 & 1 \end{bmatrix} \) is the basic polarization kernel. The \( \otimes n \) denotes the \( n \)-fold Kronecker power \cite{liaoContrucofpolarCode2022}.

Decoding of polar codes typically uses Successive Cancellation (SC), where the likelihood of each bit is computed sequentially based on previously decoded bits. The Log-Likelihood Ratio (LLR) update for bit \( u_i \) is given by,

\begin{equation}
L(u_i) = \log \frac{P(y|u_1, \ldots, u_{i-1}, u_i=0)}{P(y|u_1, \ldots, u_{i-1}, u_i=1)}.
\end{equation}

To enhance performance in noisy environments, Successive Cancellation List (SCL) decoding is used, where multiple decoding paths are explored, and the most likely paths are retained \cite{Aliieeeaccess2024}. The inclusion of a Cyclic Redundancy Check (CRC) further aids in path selection, forming the CRC-Aided SCL (CA-SCL) method \cite{bioglio2020design}. Additionally, BP decoding has been adapted for polar codes, providing a parallel decoding alternative that updates the LLRs iteratively, similar to LDPC decoding.

\subsection{Various ML Approaches for Channel Coding}
ML is playing an increasingly important role in channel coding, helping to enhance encoding, decoding, and autoencoding processes. By addressing the limitations of traditional methods, ML introduces new ways to optimize performance in complex communication systems. We categorize ML techniques in channel coding into two main areas: RL and DL, as summarized in Table~\ref{tab:ml_techniques}. In the following subsections, we focus on decoding techniques based on these methods.

\subsubsection{RL}
\begin{figure*}[htbp!]
\includegraphics[width=0.6\linewidth]{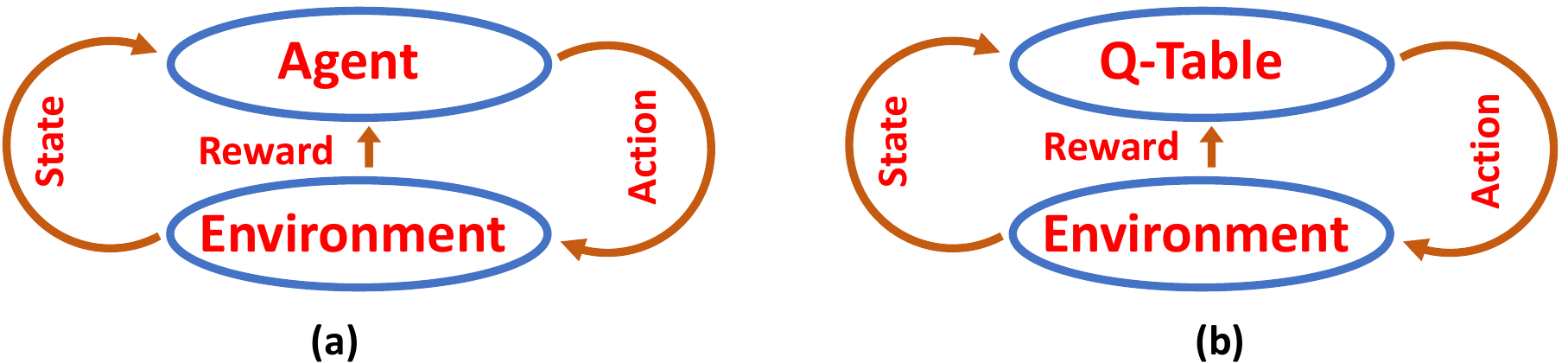}
\centering
\caption{(a) Standard RL framework where the agent interacts with the environment by taking actions and receiving states and rewards. (b) Q-Learning framework where a Q-table is used to store and update the optimal action-value pairs based on interactions with the environment.}
\label{Fig11}
\end{figure*}

RL is a form of ML in which an agent adapts through learning to accomplish predefined objectives by interacting with its environment \cite{liaoContrucofpolarCode2022, jang2021improving, HuangAICoding2020}. The agent's actions result in rewards or penalties, and its goal is to optimize the total rewards over time. RL is typically represented using the Markov Decision Process (MDP), which models decision-making through a framework of states, actions, and rewards, as depicted in Fig.~\ref{Fig11}a.

In RL, rewards are typically binary, with a ``1'' indicating a successful action and a ``-1'' representing an unsuccessful one. The agent tries to refine its decision-making policy to maximize long-term rewards. There are various approaches to learning these policies in RL: value-driven approaches, like Q-learning, which approximate the value of each action in a given state; policy-based methods, which directly optimize the policy itself; and model-based methods, which involve simulating the environment and using predictions to guide decision-making \cite{wang2021reinforcement}. Examples of model-based approaches include lookahead search and trajectory optimization, where the agent plans future actions based on simulated outcomes.

The state space \( s \) includes all possible states of the system, while the action space \( a \) consists of every action the agent can take. After performing an action \( a \) in state \( s \), the agent transitions to a new state \( s' \) and receives a reward \( R \). The value-based approach, such as Q-learning \cite{wang2021reinforcement}, focuses on determining the optimal policy by estimating the value of actions in different states. The Q-learning update rule is given as follows,
\begin{equation}
Q'(s, a) = Q(s, a) + \alpha \cdot \left[ R + \gamma \cdot \max_{a'} Q(s', a') - Q(s, a) \right],
\end{equation}
where \( \alpha \) is the learning rate, and \( \gamma \) is the discount factor. Fig.~\ref{Fig11}b shows how Q-learning applies to channel decoding, where a Q-table stores reward values to guide the decoding process. A modified version of Q-learning, known as clustered Q-learning \cite{habib2020learning}, optimizes the Tanner graph in LDPC codes, improving decoding efficiency and reducing computational complexity.

RL's main strength lies in its ability to optimize long-term rewards, and it allows for custom reward systems to guide learning in more complex environments, as discussed in \cite{tung2021effective}. This flexibility is particularly useful for adapting to varying channel conditions in communication systems.
\begin{table*}[htbp!]
\centering
\caption{Categorization of ML Techniques for Channel Coding}
\label{tab:ml_techniques}
\scriptsize
\begin{tabular}{>{\raggedright\arraybackslash}p{1.5cm} >{\raggedright\arraybackslash}p{2.5cm} >{\raggedright\arraybackslash}p{3cm} >{\raggedright\arraybackslash}p{5cm}}
\toprule
\textbf{Category} & \textbf{Type} & \textbf{Subcategory} & \textbf{Description and Key Features} \\ 
\midrule
\multirow{8}{*}{\rotatebox{90}{\parbox{0.25cm}{\centering \textbf{RL}}}} 
 & \raisebox{-4mm}{Standard RL} & Basic RL Framework & 
 Trains agents to maximize rewards by interacting with an environment. Utilizes Markov Decision Process (MDP) as a modeling framework. \\ 
\cmidrule{2-4} 
 & \raisebox{-2.5mm}{Deep RL} & Combines RL and DL techniques & 
 Use neural networks to approximate policies or value functions. \\ 
\midrule
\multirow{10}{*}{\rotatebox{90}{\textbf{DL}}} 
 & Convolutional & Convolutional Neural Networks (CNN) & 
 Effective for feature extraction and image-like data processing. \\ 
\cmidrule{2-4} 
 & Recurrent & Recurrent Neural Network (RNN) & 
 Useful for sequential data processing, such as time-series. \\ 
\cmidrule{2-4} 
 & Federated & FL & 
 Collaborative ML approach where models are trained across decentralized data. \\ 
\cmidrule{2-4} 
 & Other DL Techniques & Any DL method outside CNN, RNN, or FL & 
 Includes innovative methods that cater to unique channel coding problems. \\ 
\bottomrule
\end{tabular}
\end{table*}
\subsubsection{Deep RL}
DRL has emerged as a promising approach for decoding channel codes in complex communication systems \cite{tung2021effective}. The decoding goal is to recover codewords \( \mathbf{x} \) from noisy observations \( \mathbf{r} = \mathbf{G} \mathbf{x} + \mathbf{z} \), where \( \mathbf{G} \) is the channel matrix and \( \mathbf{z} \) denotes noise. Traditional decoding methods like maximum likelihood are accurate but computationally infeasible for long codes or high-dimensional modulation. DRL models decoding as an MDP, where state \( s_t \) reflects partial decoding, action \( a_t \) selects a bit estimate or codeword, and reward \( r_t \) reflects decoding accuracy \cite{9406115}. Using algorithms like Proximal Policy Optimization (PPO) or Deep Q-Networks (DQN), the agent learns a policy \( \pi_{\theta}(a_t | s_t) \) to maximize cumulative reward \( R = \sum_{t=0}^T r_t \). The policy is optimized using the policy gradient theorem,
\begin{equation}
\nabla_\theta J(\theta) = \mathbb{E}_{\pi_\theta} \left[ \nabla_\theta \log \pi_\theta(a_t | s_t) Q^\pi(s_t, a_t) \right],
\end{equation}
where \( Q^\pi(s_t, a_t) \) is the expected return from action \( a_t \) in state \( s_t \).
Integrating Graph Neural Networks (GNNs) into DRL enhances learning by capturing structural dependencies among code bits \cite{tung2021effective}. DRL-based decoders have shown robustness to noise, adaptability to nonlinear channels, and scalability with code length and modulation complexity, outperforming traditional methods under challenging conditions.

\rc{Although RL and DRL offer adaptability in dynamic wireless environments, their practical deployment remains challenging. Classical tabular RL scales poorly to large or continuous state spaces, while DRL introduces additional training complexity, sample inefficiency, and convergence sensitivity. Moreover, reward design is highly problem-dependent, and online adaptation may incur non-negligible latency and computational overhead. These limitations suggest that RL-based methods are most suitable for scenarios requiring adaptive decision-making under model uncertainty, rather than as universal replacements for structured model-based approaches.}
\subsection{Improved Channel Coding by DL Approaches}
DL has become a powerful enabler in channel coding, offering the capacity to model nonlinear relationships and adapt to diverse noise conditions in wireless communications \cite{meenalakshmi2024deep}. Leveraging Artificial Neural Networks (ANNs), DL frameworks can learn optimal encoding and decoding functions directly from data, improving both reliability and computational efficiency.

A typical feed-forward network performs a nonlinear transformation of the input \( x \) through a sequence of layers, as,
\begin{equation}
f(x) = \varphi_k\left(W_k \cdots \varphi_1(W_1 x + b_1) \cdots + b_k \right),
\end{equation}
where \( W_i \), \( b_i \), and \( \varphi_i \) denote weights, biases, and activation functions (e.g., ReLU), respectively. These nonlinear activations are critical; without them, even deep networks would behave as linear models. For instance, in a basic two-layer network,
\begin{equation}
h = \text{ReLU}(W_1 x + b_1), \quad y = W_2 h + b_2.
\end{equation}
Training is typically achieved by minimizing a task-specific loss. For binary channel decoding, the cross-entropy loss is commonly used,
\begin{equation}
L = -\frac{1}{N} \sum_{i=1}^N \left[ y_i \log(\hat{y}_i) + (1 - y_i) \log(1 - \hat{y}_i) \right],
\end{equation}
where \( y_i \) and \( \hat{y}_i \) are the true and predicted outputs.
End-to-end learning frameworks further extend DL's utility by jointly optimizing neural encoders and decoders. Given an encoder \( \mathcal{E}_\phi \) and decoder \( \mathcal{D}_\theta \), the training objective is,
\begin{equation}
\min_{\phi, \theta} \mathbb{E}_{x, e} \left[ \|x - \mathcal{D}_\theta(\mathcal{E}_\phi(x) + e)\|^2 \right],
\end{equation}
enabling the model to adapt to channel impairments such as fading and noise \cite{xiang2023polar, mosallaei2024enhancing}.

By replacing manually designed algorithms with data-driven models, DL enhances robustness, reduces latency, and enables parallelized decoding on modern hardware. These capabilities make DL-based coding particularly attractive for next-generation communication systems, where adaptability and efficiency are critical.
\subsubsection{Using CNN}
CNNs have proven effective in channel coding, particularly under challenging conditions like noise, fading, and interference \cite{LiangNeuralEnhancedBelief2021}. Their ability to extract local features and spatial correlations enables them to improve both encoding and decoding performance \cite{9447977, liang2018iterative}. Let \( x \in \mathbb{R}^k \) denote the input data. A CNN-based encoder \( \mathcal{E}_\phi(x) \), parameterized by \( \phi \), maps \( x \) to a codeword \( c \in \mathbb{R}^n \), where \( n > k \),
\begin{equation}
    c = \mathcal{E}_\phi(x) = f(W * x + b),
\end{equation}
with \( W \) as convolution filters, \( * \) the convolution operator, \( b \) the bias, and \( f(\cdot) \) an activation function (e.g., ReLU). The encoded signal \( c \) is transmitted through a noisy channel,
\begin{equation}
    r = c + e, \quad \text{with } e \sim \mathcal{N}(0, \sigma^2 I).
\end{equation}
At the receiver, a CNN decoder \( \mathcal{D}_\theta(r) \) reconstructs the message,
\begin{equation}
    \hat{x} = \mathcal{D}_\theta(r) = g(W' * r + b'),
\end{equation}
where \( W' \), \( b' \), and \( g(\cdot) \) correspond to decoder parameters and activations.

Training typically minimizes reconstruction error using mean squared error,
\begin{equation}
    \min_{\phi, \theta} \mathbb{E}_{x, e} \left[ \|x - \mathcal{D}_\theta(\mathcal{E}_\phi(x) + e)\|^2 \right],
\end{equation}
or cross-entropy loss for binary data,
\begin{equation}
    \mathcal{L}(\phi, \theta)= -\frac{1}{N} \sum_{i=1}^N \sum_{j=1}^k \big[ x_{ij} \log \hat{x}_{ij} + (1 - x_{ij}) \log (1 - \hat{x}_{ij}) \big].
\end{equation}
The hierarchical structure of CNNs enables them to identify complex bit dependencies and spatial patterns, making them resilient to noise and multipath fading. Their capacity for real-time, parallelized processing enhances decoder efficiency, making CNNs highly applicable to high-throughput, low-latency systems in 5G and beyond.
\subsubsection{RNNs}
RNNs are well-suited for channel decoding due to their ability to model sequential data and temporal dependencies, properties inherent in many error-correcting codes. In this context, the received noisy sequence \( \mathbf{r} = \mathbf{G} \mathbf{x} + \mathbf{z} \), with \( \mathbf{x} \) as the transmitted codeword, \( \mathbf{G} \) the channel matrix, and \( \mathbf{z} \) the additive noise, is treated as a time series. Unlike traditional algorithms (e.g., Viterbi or belief propagation), RNNs learn decoding rules directly from data \cite{davey2022using}.

Given an input sequence \( \mathbf{r} = \{r_1, r_2, \dots, r_T\} \), the RNN updates its hidden state \( \mathbf{h}_t \) at each time step \( t \) using,
\begin{equation}
    \mathbf{h}_t = f_\theta(\mathbf{h}_{t-1}, r_t),
\end{equation}
where \( f_\theta \) is the RNN cell (e.g., standard, LSTM, or GRU), parameterized by \( \theta \). The decoder output is computed as,
\begin{equation}
    \hat{x}_t = g_\phi(\mathbf{h}_t),
\end{equation}
with \( g_\phi \) typically a fully connected layer followed by softmax activation.

Training involves minimizing cross-entropy loss between predicted and actual sequences,
\begin{equation}
    \mathcal{L} = -\sum_{t=1}^{T} \sum_{k=1}^{K} x_t^{(k)} \log \hat{x}_t^{(k)},
\end{equation}
where \( x_t^{(k)} \) and \( \hat{x}_t^{(k)} \) are true and predicted probabilities over \( K \) output symbols.

RNN-based decoders are particularly effective for structured codes like convolutional codes, where sequence history is vital \cite{moller2024efficient, kanzarkar2021introduction}. Bidirectional RNNs further improve accuracy by incorporating both past and future context. Moreover, integrating attention mechanisms enables the model to focus on the most informative parts of the input, enhancing performance under severe noise or fading conditions. These capabilities position RNNs as robust alternatives to classical decoders in modern communication systems.
\subsubsection{FL}
FL enables distributed model training across edge devices while preserving data privacy by avoiding raw data exchange \cite{su2021secure}. Each device updates a local model using its private dataset and sends model updates \( \Delta \mathbf{w}_i \) to a central server, which aggregates them via weighted averaging,
\begin{equation}
    \mathbf{w}_{t+1} = \sum_{i=1}^N \frac{n_i}{n} \Delta \mathbf{w}_i,
\end{equation}
where \( n_i \) is the number of local samples on device \( i \), and \( n = \sum_{i=1}^N n_i \). Each device minimizes its local loss,
\begin{equation}
    \mathcal{L}_i(\mathbf{w}) = \frac{1}{n_i} \sum_{j=1}^{n_i} \ell(\mathbf{w}; \mathbf{x}_j, y_j),
\end{equation}
leading to a global optimization objective,
\begin{equation}
    \min_{\mathbf{w}} \mathcal{L}(\mathbf{w}) = \sum_{i=1}^N \frac{n_i}{n} \mathcal{L}_i(\mathbf{w}).
\end{equation}
To reduce communication overhead, FL transmits only model updates instead of raw data. Compression techniques such as quantization reduce bandwidth usage, $\tilde{\Delta \mathbf{w}}_i = \text{Quantize}(\Delta \mathbf{w}_i) + \mathbf{e}_i$, where \( \mathbf{e}_i \) is quantization noise. Sparsification methods retain only the most significant updates, $\Delta \mathbf{w}_i^{\text{sparse}} = \mathcal{S}(\Delta \mathbf{w}_i)$, with \( \mathcal{S}(\cdot) \) zeroing out components below a threshold.

FL also enhances privacy via differential privacy, adding calibrated Gaussian noise,
\begin{equation}
    \Delta \mathbf{w}_i' = \Delta \mathbf{w}_i + \mathbf{n}, \quad \mathbf{n} \sim \mathcal{N}(0, \sigma^2 \mathbf{I}),
\end{equation}
where \( \sigma \) controls the privacy-accuracy trade-off.

In communication systems, model updates may be distorted by channel noise,
\begin{equation}
    \mathbf{r} = \mathbf{G} \Delta \mathbf{w}_i + \mathbf{z},
\end{equation}
where \( \mathbf{G} \) is the channel matrix and \( \mathbf{z} \) denotes additive noise. FL remains robust in such conditions by leveraging its decentralized, communication-efficient structure. It is particularly advantageous in edge environments with dynamic connectivity, offering low-latency, private, and adaptive learning aligned with the demands of next-generation wireless systems.
\subsection{Neural BP for Improved LDPC Decoding}
Neural network-optimized BP decoders represent a transformative advancement in the decoding of error-correcting codes, particularly LDPC codes. Traditional BP algorithms operate on the Tanner graph representation of the code structure, where messages LLRs are exchanged iteratively between variable nodes and check nodes  \cite{9406115}. The objective is to refine the estimation of the transmitted codeword by solving a system of probabilistic equations defined by the parity-check matrix \( \mathbf{\bm H} \), ensuring that \( \mathbf{\bm H} \cdot \mathbf{\bm c}^T = \mathbf{0} \mod 2 \), where \( \mathbf{\bm c} \) represents the codeword vector \cite{mosallaei2024enhancing}. In the conventional BP, the message updates for a variable node \( v_i \) and a check node \( c_j \) are governed by the following update rules,
\begin{equation}
    m_{v_i \to c_j} = L(v_i) + \sum_{c_k \in \mathcal{N}(v_i) \setminus c_j} m_{c_k \to v_i},
\end{equation}
and,
\begin{equation}
    m_{c_j \to v_i} = 2 \tanh^{-1} \left( \prod_{v_k \in \mathcal{N}(c_j) \setminus v_i} \tanh\left(\frac{m_{v_k \to c_j}}{2}\right) \right),
\end{equation}
where \( L(v_i) \) is the channel LLR for variable node \( v_i \), \( m_{v_i \to c_j} \) is the message passed from node \( v_i \) to check node \( c_j \), and \( \mathcal{N}(v_i) \) represents the set of neighbors of \( v_i \) in the Tanner graph. Despite BP's effectiveness, it often suffers from slow convergence, high computational complexity, and degraded performance under high-noise or correlated channel conditions.

Neural network-enhanced BP decoders mitigate these issues by introducing trainable parameters into the message-passing process. This is typically achieved through techniques such as deep unfolding, where the iterative BP decoding process is unrolled into a fixed number of layers, with each layer modeled as a neural network with learnable weights \cite{shi2022deep}. The message update rules are thus modified as,
\begin{equation}
    m_{v_i \to c_j} = f_{\theta_1^{(t)}} \left( L(v_i), \sum_{c_k \in \mathcal{N}(v_i) \setminus c_j} m_{c_k \to v_i} \right),
\end{equation}
and,
\begin{equation}
    m_{c_j \to v_i} = g_{\theta_2^{(t)}} \left( \prod_{v_k \in \mathcal{N}(c_j) \setminus v_i} \tanh\left( \frac{m_{v_k \to c_j}}{2} \right) \right),
\end{equation}
where \( f_{\theta_1^{(t)}} \) and \( g_{\theta_2^{(t)}} \) are neural network-based functions parameterized by weights \( \theta_1^{(t)} \) and \( \theta_2^{(t)} \) at the \( t \)-th iteration (or layer). These weights are learned during training to minimize a loss function, commonly the BER or Block Error Rate (BLER), evaluated under various channel conditions. 

By optimizing these parameters, the neural network-enhanced BP decoder can adapt the LLR update rules dynamically to better suit the noise characteristics and channel structure \cite{mosallaei2024enhancing}. For instance, in scenarios with correlated noise, the trainable parameters allow the decoder to learn a correction mechanism, which would be difficult to model analytically using traditional BP. Neural network-enhanced BP decoders combine the interpretability of classical BP with the adaptability of DL, resulting in lower error rates and faster convergence \cite{yang2024hidden}. This method greatly enhances the performance of LDPC decoding, especially in challenging communication environments with high noise levels or complex channel impairments.
\subsubsection{Optimizing Channel Decoding with Neural Processing}
Neural post-processing techniques for channel decoders offer an innovative approach to enhancing decoding performance by augmenting traditional decoders with neural networks. These techniques refine the output of the decoder and provide corrective feedback for further iterative refinement, particularly in environments with correlated noise or complex channel conditions.

\begin{figure*}[ht!]
\includegraphics[width=0.5\linewidth]{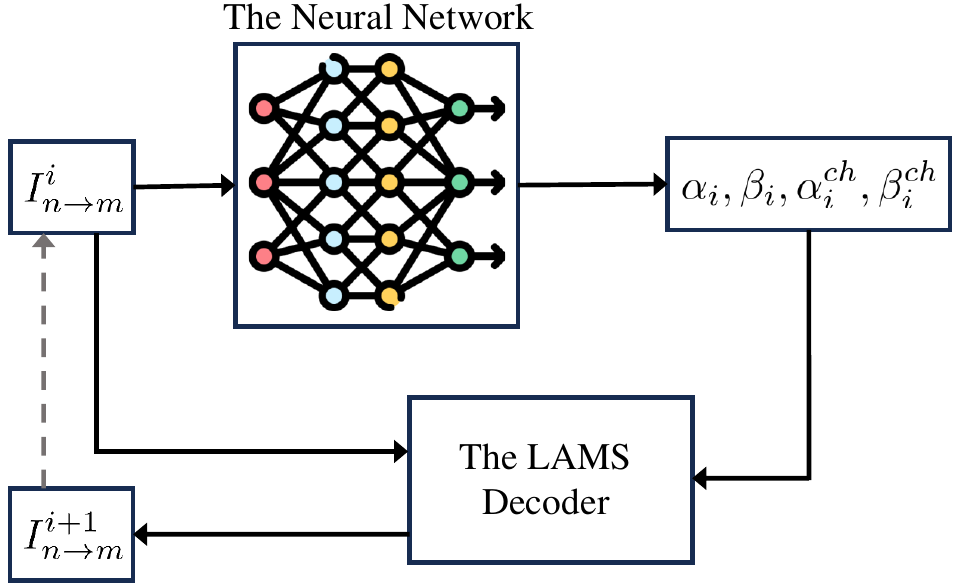}
\centering
\caption{Neural network-assisted LAMS decoder for refining normalization and offset parameters in BP decoding, \rc{adapted from \cite{wu2018decoding}.}}
\label{Fig3N}
\end{figure*}

For this purpose, Fig.~\ref{Fig3N} illustrates a neural network-based approach for enhancing normalization and offset parameters in the Linear Approximation Min-Sum (LAMS) decoder \cite{wu2018decoding}. While this method introduces a neural network to improve the decoder's performance, it primarily focuses on parameter optimization within the LAMS framework. A key limitation of this approach is its reliance on predefined normalization and offset parameters, which may not fully capture the complex noise patterns present in real-world communication channels. This restricts the decoder's ability to adapt to highly correlated or non-Gaussian noise, limiting its effectiveness in challenging environments.

\begin{figure*}[ht!]
\includegraphics[width=0.8\linewidth]{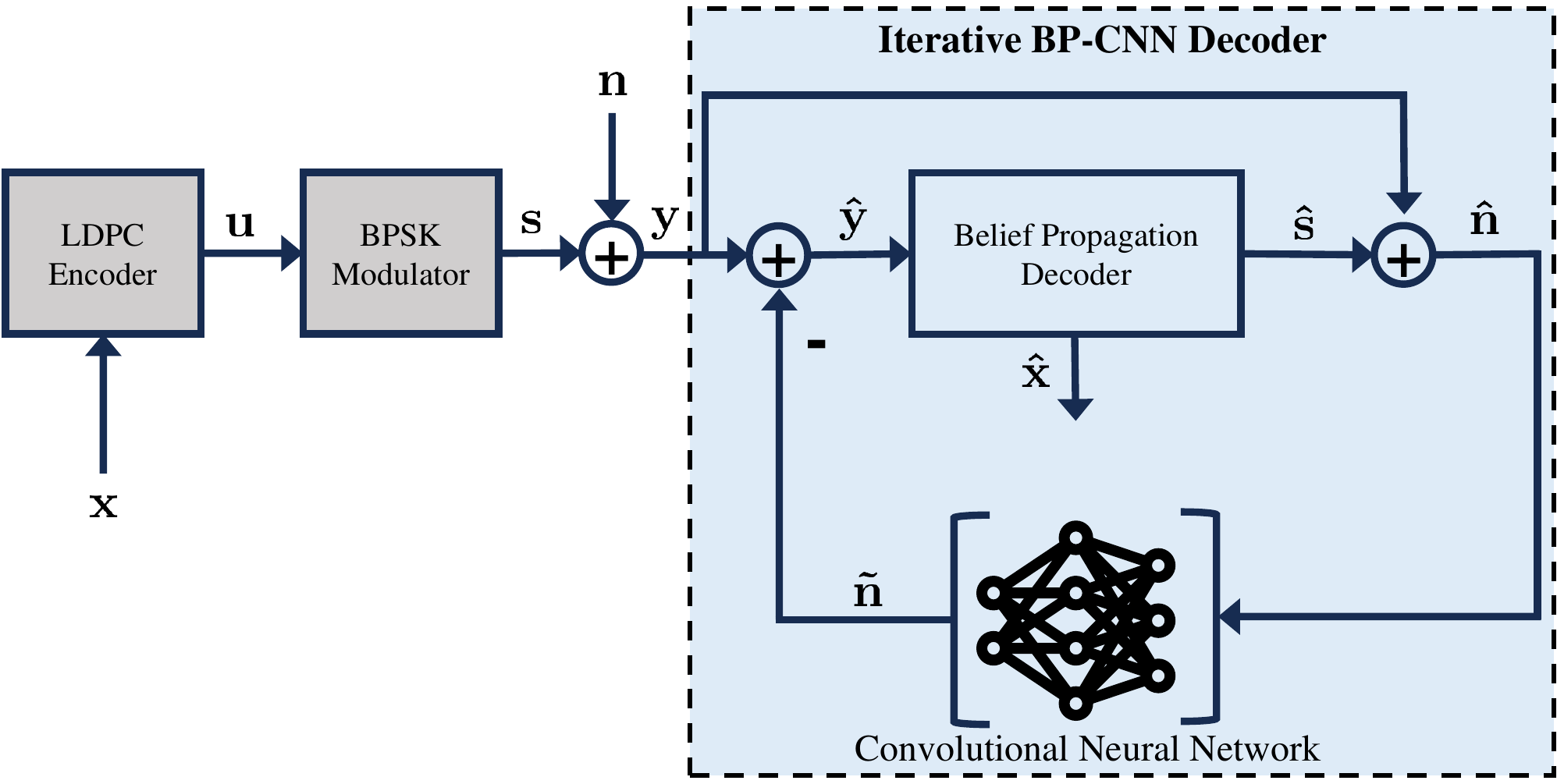}
\centering
\caption{Iterative BP-CNN decoder integrating a CNN for enhanced noise estimation and decoding accuracy, \rc{adapted from \cite{liang2018iterative}.}}
\label{Fig5}
\end{figure*}

To address these limitations, a more advanced approach that combines CNNs with iterative BP decoding is proposed in \cite{liang2018iterative}, which is illustrated in Fig.~\ref{Fig5}. In this BP-CNN architecture, the BP decoder first estimates the transmitted symbols \( \mathbf{\hat{s}} \). The estimated symbols are subtracted from the received signal \( \mathbf{y} \), yielding an estimate of the noise, \( \mathbf{\hat{n}} = \mathbf{y} - \mathbf{\hat{s}} \). The CNN then processes this estimated noise \( \mathbf{\hat{n}} \), refining it to minimize the residual estimation error. The noise model can be represented as \( \mathbf{\hat{n}} = \mathbf{n} + \xi \), where \( \mathbf{n} \) represents the true noise, and \( \xi \) is the estimation error. The CNN’s refined noise \( \mathbf{\tilde{n}} \) is then subtracted from \( \mathbf{y} \), producing an updated signal for further decoding,
\begin{equation}
    \mathbf{\hat{y}} = \mathbf{y} - \mathbf{\tilde{n}} = \mathbf{s} + \mathbf{r},
\end{equation}
where \( \mathbf{r} \) is the residual noise. This iterative process leverages CNN's ability to model complex noise patterns, such as channel correlations, significantly enhancing decoding accuracy and SNR.
To optimize the CNN training, a specialized loss function is used,
\begin{equation}
    \text{Loss}_B = \frac{\|\mathbf{r}\|^2}{N} + \lambda \left(S^2 + \frac{1}{4}(C - 3)^2\right),
\end{equation}
where \( \frac{\|\mathbf{r}\|^2}{N} \) minimizes the power of the residual noise \( \mathbf{r} \), \( S \) and \( C \) are the skewness and kurtosis from the Jarque-Bera normality test, and \( \lambda \) is a regularization parameter. The term \( S^2 + \frac{1}{4}(C - 3)^2 \) encourages the residual noise to approximate a Gaussian distribution, aligning with the assumptions of BP decoding. Simulation results demonstrate that this framework achieves significant decoding improvements under correlated noise, particularly as the correlation strength \( \eta \) increases. The system is also robust to mismatched conditions and can adapt to diverse noise models through re-training, highlighting its practical application in real-world communication systems.

Another effective neural post-processing approach integrates LSTM networks to refine the decoding of polar codes \cite{meenalakshmi2024deep}. In scenarios where a CRC fails, the LSTM identifies and corrects the first erroneous bit in the output of a Successive Cancellation (SC) decoder, significantly enhancing subsequent decoding attempts. The LSTM refines the SC decoder output \( \hat{c} \) as,
\begin{equation}
    \mathbf{\hat{c}'} = \text{LSTM}(\mathbf{\hat{c}}),
\end{equation}
where \( \mathbf{\hat{c}} \) is the SC decoder output, and \( \mathbf{\hat{c}'} \) is the corrected codeword candidate. This LSTM-assisted SC flip algorithm outperforms the CRC-aided SCL algorithm while maintaining a limited number of decoding attempts, thus remaining computationally efficient. 

Additionally, LSTMs have been integrated into SCL decoding to enhance bit-flipping strategies. The LSTM dynamically predicts which bits are most likely to be erroneous, improving the bit-flipping process,
\begin{equation}
    \text{Flip} = \text{LSTM}(\mathbf{P}_{\text{log}}),
\end{equation}
where \( \mathbf{P}_{\text{log}} \) represents the log-likelihood probabilities of the decoded bits. This approach achieves state-of-the-art decoding performance while keeping computational resources constrained.

Neural post-processing techniques are modular and independent of the primary decoding algorithm, providing a flexible and adaptive means of enhancing error correction. By iteratively refining the decoder’s output through neural network feedback, these techniques address residual errors and mitigate the effects of correlated noise, making them a valuable tool for robust and efficient decoding in modern communication systems.
\subsection{Road Map for AI-Driven Optimization of Channel Code Construction}  
LDPC codes are characterized by a sparse parity-check matrix \( \mathbf{H} \), where the codewords \( \mathbf{c} \) satisfy the condition \( \mathbf{H} \cdot \mathbf{c} = 0 \mod 2 \) \cite{mosallaei2024enhancing}. The sparsity of \( \mathbf{H} \) enables efficient decoding, especially when combined with iterative algorithms such as BP. Optimizing LDPC codes involves searching for matrices that minimize the BER while maintaining decoding efficiency. DRL can be employed to train neural networks for iterative code design, with the objective of reducing BER. The associated loss function is defined as,
\begin{equation}
    L = \text{BER}_{\text{AI-designed}} - \text{BER}_{\text{optimal}},
\end{equation}
where \( \text{BER}_{\text{optimal}} \) represents the best achievable BER for a given system. To guide neural networks toward configurations that approximate the performance of Progressive Edge Growth (PEG) designs, Monte Carlo Tree Search (MCTS) is utilized. Additionally, Genetic Algorithms (GA) evolve populations of codes by optimizing performance metrics such as BER or decoding iterations, leading to improved performance over standard 5G LDPC codes \cite{alhijawi2024genetic}. The GA optimization process aims to find the optimal code \( \mathbf{C}^{'} \) by minimizing the BER, expressed as,
\begin{equation}
        \mathbf{C}^{'} = \arg \min_{\mathbf{C}} \text{BER}(\mathbf{C}),
\end{equation}
where \( \mathbf{C}^{'} \) represents the optimized LDPC code.
Polar codes, which rely on the concept of channel polarization, transform binary channels into polarized ones. The encoding process involves selecting reliable channels for transmitting information bits while freezing the less reliable ones. This can be mathematically expressed as:
\[
    \mathbf{x} = \mathcal{G}_N(\mathbf{u}, \mathbf{f}),
\]
where \( \mathcal{G}_N \) is the polarizing transform matrix, \( \mathbf{u} \) denotes the information bits, and \( \mathbf{f} \) represents the frozen bits. AI techniques are applied to optimize the selection of frozen bits for decoding algorithms such as SC, BP, and SCL to minimize the BER. This can be framed as,
\begin{equation}
        \mathbf{f}^{'} = \arg \min_{\mathbf{f}} \text{BER}(\mathbf{f}, \mathbf{u}),
\end{equation}
where \( \mathbf{f}^{'} \) denotes the optimized frozen bit selection.

Neural networks treat bit assignments as trainable parameters, adjusting them iteratively to minimize BER under varying channel conditions, such as AWGN or Rayleigh fading. The associated loss function is defined as,

\begin{equation}
    \mathcal{L} = \text{BER}_{\text{trained}} - \text{BER}_{\text{baseline}},
\end{equation}
where \( \text{BER}_{\text{trained}} \) refers to the BER obtained after training, and \( \text{BER}_{\text{baseline}} \) represents the performance of the baseline decoder.

AI-based optimization methods can be categorized into model-driven approaches, which leverage domain-specific knowledge, and data-driven approaches, which use algorithms like DRL and neural networks to optimize code design. These techniques dynamically adapt encoding and decoding strategies to match varying channel conditions, thus enhancing error correction performance while managing hardware complexity for practical deployment in real-world communication systems.

\rc{\subsection{Critical Analysis of DL-Based Channel Coding}
While DL-based channel coding techniques have demonstrated promising performance in several studies, their advantages over classical coding schemes, such as LDPC and Turbo codes, are highly context-dependent \cite{MatsumineOchiai_OJCOMS_2024_DLChannelCodingSurvey}. DL-based decoders tend to outperform classical methods in scenarios involving short block lengths \cite{paper8ShortBlocklength2023,dorner2022learning}, non-linear channel impairments \cite{freire2023reducing,ye2017initial}, or when channel models deviate from ideal assumptions \cite{dorner2022learning}. In such cases, neural decoders can learn implicit structures and compensate for model mismatches more effectively than traditional algorithms. However, for long block lengths and well-characterized channels (e.g., AWGN), classical coding schemes remain highly optimized and often outperform DL-based approaches in both reliability and computational efficiency \cite{costello2007channel}. Furthermore, neural decoders typically require extensive training and may not generalize well across varying channel conditions. These observations suggest that DL-based channel coding is not a universal replacement but rather a complementary approach that is particularly beneficial in complex or poorly modeled environments \cite{MatsumineOchiai_OJCOMS_2024_DLChannelCodingSurvey}.}

\rc{\subsection{Limitations and Failure Modes of Neural Decoders}
Despite their flexibility, neural decoders exhibit several limitations. One key challenge is their sensitivity to distribution shifts; models trained under specific channel conditions may experience significant performance degradation when deployed in environments with different distributions \cite{ dai2023toward, lyu2018performance}. Additionally, neural decoders often require large amounts of labeled training data, which may not be feasible in practical communication systems. Overfitting is another concern, particularly for complex architectures trained on limited datasets \cite{zhi2023multi}. Computational complexity and inference latency also remain significant barriers, especially for real-time applications \cite{freire2023reducing}. Unlike classical decoders with well-defined iterative structures, neural decoders may involve deep architectures that are less interpretable and harder to optimize for hardware implementation \cite{liang2018iterative}. Additionally, the robustness to rare or adversarial noise patterns remains an open challenge, underscoring the need for more reliable and generalizable DL-based decoding strategies.}

\rc{\subsection{Discussion of Contradictory Findings}
The existing body of research presents mixed and sometimes contradictory findings regarding the effectiveness of DL-based channel coding techniques. While several studies report noticeable performance gains, particularly in short block-length regimes or under non-ideal channel conditions, other works demonstrate that these advantages diminish or disappear in standardized benchmark scenarios such as AWGN channels with long block lengths. These discrepancies can largely be attributed to differences in experimental setups, including variations in channel models, SNR ranges, training methodologies, and evaluation metrics. In many cases, reported improvements are highly specific to the conditions under which the models were trained and evaluated. Additionally, some studies emphasize gains in specific metrics such as BER, while others focus on perceptual or task-oriented metrics, further complicating direct comparison. As a result, it is challenging to draw universally applicable conclusions regarding the superiority of DL-based approaches. This inconsistency underscores the importance of developing standardized evaluation frameworks and benchmarking methodologies to enable fair and reproducible comparisons across different approaches. Frameworks such as DeepMIMO \cite{hu2023understanding} and NVIDIA Sionna \cite{cammerer2025sionna} have recently been introduced to enable reproducible and standardized benchmarking of AI-based communication systems under realistic channel conditions. To provide a structured overview, Table~\ref{dl_vs_classical_coding} summarizes representative findings from selected studies, highlighting the conditions under which DL-based channel coding techniques demonstrate advantages.}
\begin{table*}[t]
\centering
\caption{Comparative Analysis of DL-Based and Classical Channel Coding Approaches}
\label{dl_vs_classical_coding}
\renewcommand{\arraystretch}{1.2}

\resizebox{\textwidth}{!}{
\begin{tabular}{cccccc}
\toprule
\textcolor{black}{\textbf{Study}} & \textcolor{black}{\textbf{Approach}} & \textcolor{black}{\textbf{Channel Model}} & \textcolor{black}{\textbf{Block Length}} & \textcolor{black}{\textbf{Reported Performance}} & \textcolor{black}{\textbf{Key Observation}} \\
\midrule

\textcolor{black}{\cite{jiang2020learn}} & \textcolor{black}{Neural Decoder} & \textcolor{black}{AWGN} & \textcolor{black}{Short} & \textcolor{black}{$\sim$0.5 dB gain} & \textcolor{black}{Effective in short block regime} \\
\midrule

\textcolor{black}{\cite{jiang2019deepturbo}} & \textcolor{black}{DeepTurbo} & \textcolor{black}{AWGN} & \textcolor{black}{Long} & \textcolor{black}{Comparable to classical} & \textcolor{black}{No consistent improvement} \\
\midrule

\textcolor{black}{\cite{njoku2021bler}} & \textcolor{black}{Autoencoder-based Coding} & \textcolor{black}{Non-linear Channel} & \textcolor{black}{Short} & \textcolor{black}{$\sim$1--1.5 dB gain} & \textcolor{black}{Robust under model mismatch} \\
\midrule

\textcolor{black}{\cite{habib2023reldec}} & \textcolor{black}{Learned LDPC Decoding} & \textcolor{black}{Fading Channel} & \textcolor{black}{Medium} & \textcolor{black}{Slight improvement} & \textcolor{black}{Gains depend on channel variability} \\
\bottomrule
\end{tabular}
}

\vspace{2mm}
\begin{minipage}{\textwidth}
\raggedright
\scriptsize{\textcolor{black}{\textit{Note:} The reported performance gains are derived from individual studies conducted under different experimental conditions and should be interpreted as indicative trends rather than directly comparable results.}}
\end{minipage}

\end{table*}
\section{Other AI-Enhanced Baseband Components}\label{otherAI-EnhancedBaseband}
In addition to the core signal processing elements, specifically the error correction decoders and MIMO detectors, developing and deploying other baseband modules like the channel estimator, equalizer, and NOMA detectors have also taken advantage of AI techniques. This utilization aims to use specific patterns, particularly nonlinear behaviors, which are often challenging for traditional methods to address. Ongoing research concentrates on advancements in algorithmic approaches and hardware realizations for these modules.
\begin{figure*}[htbp!]
\includegraphics[width=0.6\linewidth]{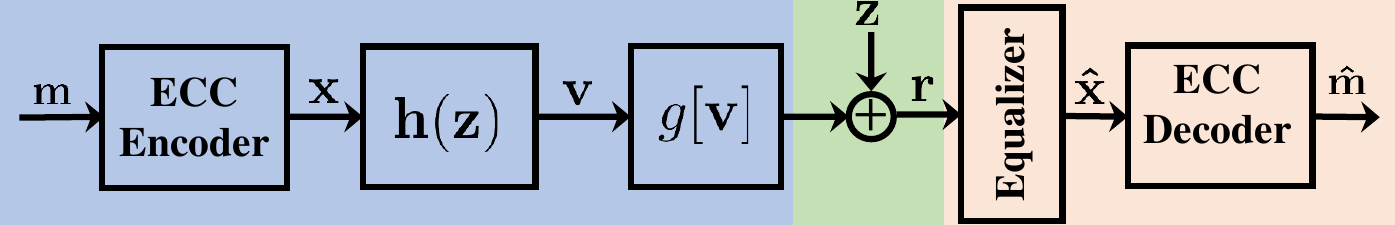}
\centering
\caption{Communication system model with ECC encoding/decoding, channel equalization, and noise representation.}
\label{Fig6}
\end{figure*}
\subsection{Equalization}
\subsubsection{Introduction}  
Channel equalization mitigates distortions from inter-symbol interference and nonlinear effects caused by amplifiers and converters \cite{moghaddam2023statistical}, as illustrated in Fig.~\ref{Fig6}. Given modulated symbols \( \mathbf{x} \) and received signal \( \mathbf{r} \), the channel can be modeled as,  
\begin{align*}
    \mathbf{v} &= \mathbf{x} \otimes \mathbf{h(z)}, \\
    \mathbf{r} &= g[\mathbf{v}] + \mathbf{z},
\end{align*}  
where \( \mathbf{h(z)} \) are channel coefficients, \( \mathbf{z} \) is AWGN, \( g \) is a nonlinear function, and \( \otimes \) denotes convolution.

NN-based equalizers have surpassed traditional linear ones in performance \cite{freire2023reducing, dorner2022learning, yang2024hidden}. However, they often alter noise characteristics, violating the Gaussian assumptions required by standard decoders and degrading performance \cite{dorner2022learning}. Iterative joint equalization and decoding methods address this, but are computationally intensive \cite{yang2024hidden}.

\subsubsection{DL-Driven Combined Equalization and Decoding}  
To improve efficiency, an end-to-end DNN approach was proposed in \cite{ye2017initial}, jointly handling equalization and decoding. It maps received signals to estimated codewords and shows gains over conventional methods for small polar codes. However, larger codes require deeper networks and longer training. Using DNNs for signal recovery introduces challenges: their fixed structure lacks flexibility for varying input lengths, and their inference cost grows with sequence size. CNNs \cite{tian2022heterogeneous} and RNNs \cite{mienye2024recurrent} address these limitations. CNNs are well-suited for ISI due to their ability to extract local features from consecutive symbols, while RNNs adapt to input variability with fewer parameters.

A hybrid CNN equalizer and DNN decoder in \cite{costa2023cnn} outperforms purely DNN-based designs. Similarly, \cite{fischer2021wiener} presents an RNN-based three-stage architecture covering equalization, LLR computation, and error correction, achieving improved accuracy and reduced complexity via gated units. These architectures offer better adaptability and efficiency than dense DNNs, supporting longer sequences and reducing computational overhead in DL-based equalization systems.

Additionally, Nonorthogonal Multiple Access (NOMA) techniques enhance spectral efficiency. Examples include Sparse-Code Multiple Access (SCMA) \cite{rebhi2021sparse}, Multi-User Shared Access (MUSA) \cite{sivalingam2021deep}, Pattern-Division Multiple Access (PDMA) \cite{deng2022hdma}, and Interleave-Division Multiple Access (IDMA) \cite{kong2022fixed}. SCMA stands out with low decoding complexity and effective spectral shaping, especially in high-dimensional constellations \cite{luo2023enhancing}.
\begin{table*}[htbp!]
\centering
\caption{AI-Based Techniques for NOMA Systems}
\label{table:noma-techniques}
\scriptsize
\begin{tabular}{@{}p{0.7cm}p{4.5cm}p{3cm}p{4.5cm}@{}}
\toprule
\multirow{21}{*}{\rotatebox{90}{\textbf{Non-Orthogonal Multiple Access}}} 
 & \textbf{Subcategory} & \textbf{Details} & \textbf{Description} \\ \cmidrule{2-4}
 & \multirow{10}{*}{Sparse-Code Multiple Access} & Codebook Construction & SCMA assigns sparse codebooks to users for efficient signal separation. \\ \cmidrule{3-4}
 &  & Detection Algorithm & Algorithms used to detect signals in a multi-user SCMA system. \\ \cmidrule{3-4}
 &  & DL & AI models for optimizing SCMA performance and codebook designs. \\ \cmidrule{3-4}
 &  & Image Processing-Based & Image-based AI methods for improving SCMA detection accuracy. \\ \cmidrule{2-4}
 & Multi-User Shared Access & Sequence-Based Separation & MUSA separates users based on unique sequence patterns for shared access. \\ \cmidrule{2-4}
 & Pattern-Division Multiple Access & AI-Based Algorithms & PDMA uses predefined patterns to allocate resources among users. \\ \cmidrule{2-4}
 & Interleave-Division Multiple Access & Data Interleaving & IDMA interleaves user data streams to achieve better separation. \\ \bottomrule
\end{tabular}
\end{table*}
\subsection{Non-Orthogonal Multiple Access}
DL offers effective solutions for decoding challenges in NOMA systems, particularly in SCMA. In SCMA, multiple users share the same resources using sparse codebooks, with each user’s symbol \( \mathbf{s}_u \) mapped to a codeword \( \mathbf{c}_u \in \mathcal{C}_u \). The transmitted signal is,

\[
\mathbf{x} = \sum_{u=1}^{U} \mathbf{s}_u \mathbf{c}_u,
\]
where \( U \) is the number of users. The main challenge lies in separating non-orthogonal signals, which traditional decoders handle with high complexity.

DL-based decoders learn the mapping from received signal \( \mathbf{y} \) to user symbols \( \mathbf{s}_u \), minimizing a loss function,

\begin{equation}
\hat{\mathbf{s}}_u = \arg \min_{\mathbf{s}_u} \mathcal{L}(\mathbf{y}, \hat{\mathbf{s}}_u),
\end{equation}
where \( \mathcal{L} \) quantifies the error (e.g., MSE or cross-entropy). Training on large datasets enables the model to decode reliably, even under noise and interference.

CNNs and RNNs are often used to exploit spatial and temporal patterns in \( \mathbf{y} \). CNNs extract hierarchical features, while RNNs capture temporal dependencies, improving decoding in dynamic channels. Training involves backpropagation and stochastic gradient descent to minimize prediction loss.

DL also supports joint estimation of the channel matrix \( \mathbf{G} \) and user symbols. Autoencoder-based architectures achieve this by optimizing,

\begin{equation}
\mathcal{L}_{AE} = \left\| \mathbf{y} - \mathbf{G} \mathbf{s} \right\|^2 + \lambda \left\| \mathbf{G} \right\|_F^2,
\end{equation}
where \( \|\cdot\|_F \) is the Frobenius norm and \( \lambda \) regularizes channel complexity. This approach enables end-to-end learning of codebook mappings and channel characteristics, improving detection in non-orthogonal settings.

Table~\ref{table:noma-techniques} summarizes AI-driven techniques for NOMA systems, including SCMA, MUSA, PDMA, and IDMA, highlighting key methods such as DL-based detection and AI-assisted sequence separation.

\subsubsection{Limitations of Existing Approaches}

In SCMA systems with \( K \) dimensions and \( M \) layers, the received signal is modeled as,

\begin{equation}
\mathbf{r} = \sum_{m=1}^M \operatorname{diag}(\mathbf{G}_m) \mathbf{x}_m + \mathbf{z},
\end{equation}

where \( \mathbf{x}_m = (x_{1m}, \ldots, x_{Km})^T \) is the input data, \( \mathbf{G}_m = (g_{1m}, \ldots, g_{Km})^T \) the channel vector, and \( \mathbf{z} \) the noise. This highlights the role of channel conditions and noise in shaping the received signal.

SCMA performance heavily relies on accurate Channel State Information (CSI). Methods like the Deterministic Message Passing Algorithm (DMPA) \cite{rebhi2021sparse} depend on precise CSI, with inaccuracies leading to performance loss. To improve detection, \cite{zhou2022channel} introduces a linear estimator for CSI and a dedicated detector to minimize interference and enhance decoding accuracy in uplink NOMA systems. However, acquiring reliable CSI remains challenging and computationally intensive. Another key limitation is codebook design. Manual approaches struggle to optimize constellation distances and structure \cite{zhang2021scma}, making efficient SCMA codebook construction a complex task that significantly impacts system performance.

\subsubsection{Blind Detection Using Image Processing}
A novel blind SCMA detector, proposed in \cite{sivalingam2021deep}, frames detection as an image processing task. It maps signal patterns into 2D images by exploiting interactions between resource and layer nodes. To handle noise, image inpainting techniques such as total variation are applied \cite{hirai2021link}. The resulting image is then denoised using a CNN, as demonstrated in \cite{ilesanmi2021methods}, where the CNN is integrated with the DMPA algorithm.

Since DMPA relies on accurate CSI, its performance degrades with poor channel estimation. The blind detection method in \cite{ilesanmi2021methods} compensates for this limitation, enhancing robustness in scenarios with CSI uncertainty.

\begin{figure*}[htbp!]
\includegraphics[width=1\linewidth]{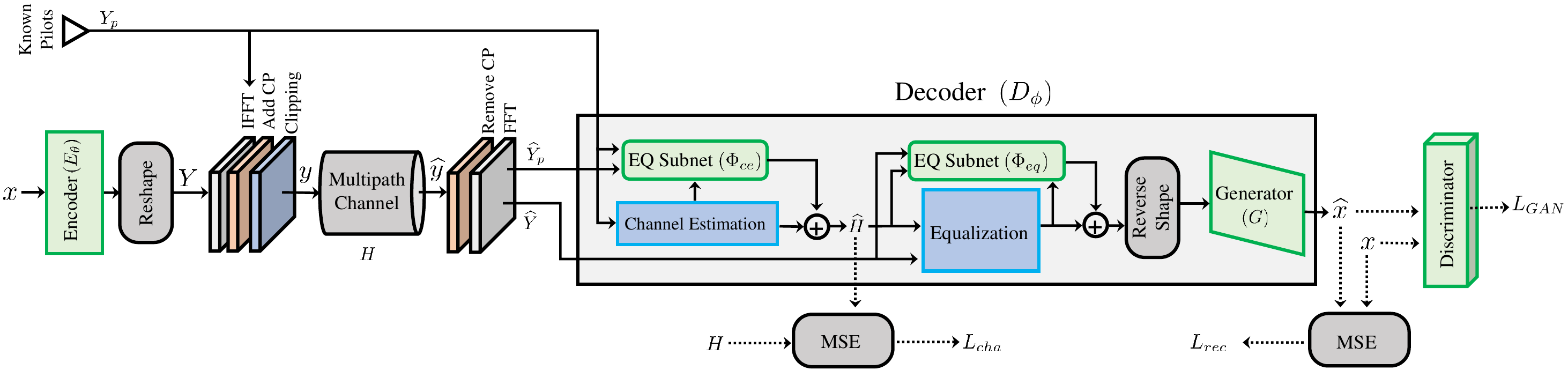}
\centering
\caption{An OFDM-guided deep JSCC framework for transmitting images over wireless networks, \rc{adapted from }\cite{paper4MingyuYang-OFDMGuidedDeepJSCC2022}. The encoder transforms input images into modulated OFDM signals, incorporating CP and optional signal clipping for peak-to-average power ratio reduction. The decoder uses domain knowledge through channel estimation, equalization, and residual subnets, while a generator reconstructs the image with adversarial loss for perceptual quality enhancement.}
\label{Fig8}
\end{figure*}

Additionally, \cite{paper4MingyuYang-OFDMGuidedDeepJSCC2022} introduces an OFDM-guided deep Joint Source Channel Coding (JSCC) framework for wireless image transmission over multipath fading channels. As shown in Fig.~\ref{Fig8}, the encoder \( E_\theta \) converts image \( x \) to frequency-domain symbols \( Y \), which are IFFT-transformed, CP-padded, and transmitted over a fading channel \( H \). The receiver uses FFT and pilot-based channel estimation (\( \hat{H} \)) followed by equalization via subnetworks \( \Phi_{eq} \) and \( \Phi_{ce} \). The decoder \( D_\phi \), aided by a generator \( G \), reconstructs the image \( \hat{x} \), while a discriminator ensures high perceptual quality using adversarial loss \( L_{GAN} \). The system jointly optimizes reconstruction (\( L_{rec} \)) and channel estimation loss (\( L_{cha} \)), enabling efficient, noise-resilient image transmission through end-to-end training.
\subsubsection{SCMA Design Based on DL}
The deep-learning-based SCMA (D-SCMA) framework proposed in \cite{kanzarkar2021introduction} leverages neural networks to autonomously construct SCMA codebooks and decoding strategies. A DNN-based decoder, \( g(\mathbf{y}; \theta_g) \), is trained to minimize the reconstruction error,
\begin{equation}
\min_{\theta_g} \left\| \mathbf{r} - g(\mathbf{y}; \theta_g) \right\|_2,
\end{equation}
where \( \mathbf{y} \) is the received signal, \( \mathbf{r} \) is the original signal, and \( \theta_g \) denotes the trainable decoder parameters. Through this optimization, the network learns data-driven detection behavior and supports adaptive SCMA codebook design. The architecture includes multiple specialized DNN units and a fully connected layer for integration, with training performed using stochastic gradient descent. Compared with manual codebook design, D-SCMA provides a more flexible and performance-oriented learning framework for SCMA detection.

Simulations reported in \cite{kanzarkar2021introduction} indicate efficiency and computational advantages over conventional methods in the studied settings, suggesting that DL is a promising tool for sparse multidimensional signal processing.

More broadly, related AI-enabled communication research has also moved toward semantic communication, where the goal is not merely to transmit raw data, but to convey task-relevant or meaningful information more efficiently \cite{yang2022semantic}. In this broader context, Table~\ref{tabcomp3} summarizes representative semantic communication advances across image, speech, and IoT applications. Additionally, the Figs.~\ref{performance_comparison} illustrate indicative \rc{trends} in PSNR, LPIPS, and BER across selected studies. Fig.~\ref{fig:ber_reduction_lollipop} summarizes the normalized BER reduction achieved by AI-based methods relative to conventional baselines, while Fig.~\ref{fig:jscc_positioning} qualitatively positions representative JSCC paradigms according to their dominant optimization objective and channel adaptability. ADJSCC emphasizes SNR-aware adaptive transmission, whereas InverseJSCC and GenerativeJSCC emphasize perceptual/semantic reconstruction quality. \rc{It is important to note that the results summarized in Figs.~\ref{performance_comparison}--\ref{fig:jscc_positioning} rescaled qualitative/normalized trend illustrations synthesized from heterogeneous studies conducted under different datasets, channel models, and experimental settings. Therefore, these figures are intended to provide qualitative and normalized insights into general performance trends rather than strict quantitative comparisons.}
\begin{figure*}[htbp!]
\includegraphics[width=0.95\linewidth]{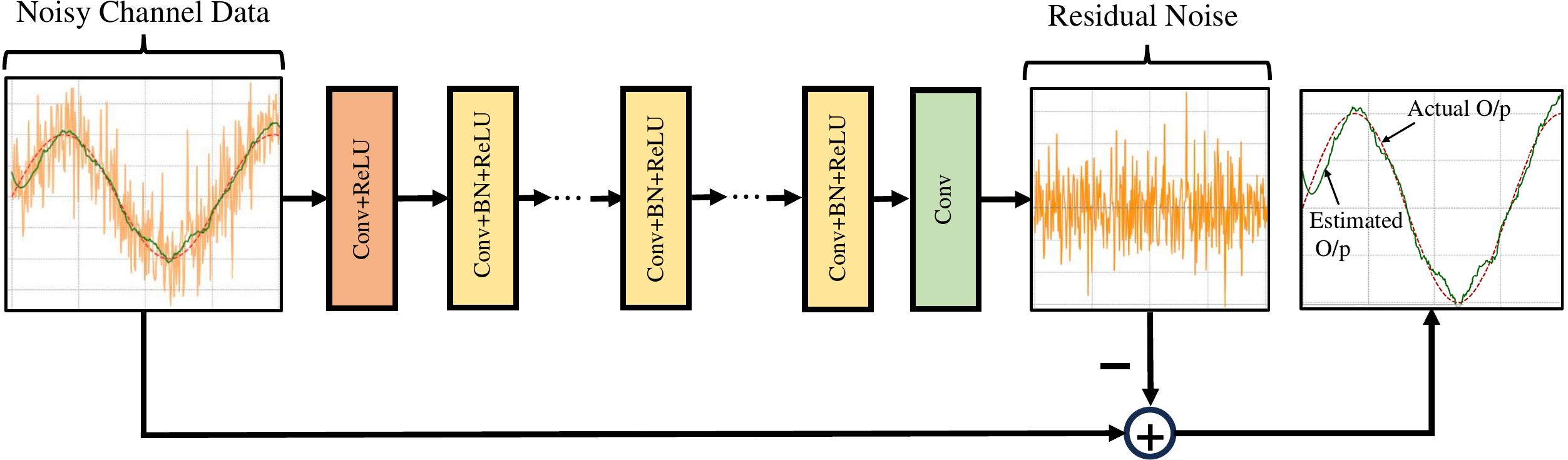}
\centering
\caption{Workflow of the DnCNN-based channel estimation, showing noisy channel input, residual noise estimation, and the reconstruction of the clean channel through convolutional layers, \rc{adapted from} \cite{paper14DeepLearning-BasedChannelEstimation2018}.}
\label{Fig7}
\end{figure*}
\subsection{Channel Estimation}
Accurate channel estimation is essential for effective detection and precoding. Traditional techniques, such as Least Squares (LS) and Linear Minimum Mean Square Error (LMMSE), are widely used LS offers simplicity but limited accuracy, while LMMSE leverages statistical information for better results. Recently, DL has advanced channel estimation by directly learning channel characteristics from data. For example, a Learned Denoising-Based Approximate Message Passing (LDAMP) network with a DnCNN denoiser outperforms compressed sensing methods \cite{wang2023versatile}. Image-based approaches treat the channel’s time-frequency response as an image, reconstructing high-resolution matrices and achieving LMMSE-level performance \cite{lv2023deep, ruan2023simplified}. These methods use DnCNN denoisers during training, as illustrated in Fig.~\ref{Fig7}, leading to improved results over traditional techniques. Furthermore, DL-based pilot design and channel estimation frameworks surpass LMMSE accuracy \cite{mashhadi2021pruning}. For rapidly changing channels, learning-based methods provide higher accuracy in dynamic conditions \cite{wang2022time}. \rc{To consolidate the above discussion and reduce purely descriptive presentation, Table~\ref{tab:compact_ai_synthesis} summarizes the main analytical contrasts between classical and AI-based approaches across the wireless communication tasks covered in this survey.}

\begin{table*}[t]
\centering
\caption{\rc{Compact analytical synthesis of AI-based techniques across key wireless communication tasks.}}
\label{tab:compact_ai_synthesis}
\tiny
\setlength{\tabcolsep}{3pt}
\renewcommand{\arraystretch}{1.08}
\begin{adjustbox}{max width=\textwidth}
\begin{tabular}{>{\RaggedRight\arraybackslash}p{1.55cm}
                >{\RaggedRight\arraybackslash}p{1.95cm}
                >{\RaggedRight\arraybackslash}p{2cm}
                >{\RaggedRight\arraybackslash}p{2.15cm}
                >{\RaggedRight\arraybackslash}p{2.15cm}
                >{\RaggedRight\arraybackslash}p{2.5cm}
                >{\RaggedRight\arraybackslash}p{1.6cm}}
\toprule
\textbf{Task} & \textbf{Classical baseline} & \textbf{AI mechanism} & \textbf{Where AI helps} & \textbf{Where AI is less clear} & \textbf{Latency / energy profile} & \textbf{Main barrier} \\
\midrule

\textbf{MIMO detection}
& ZF/MMSE, SD, AMP/OAMP, BP
& DetNet, ComNet, OAMPNet, MMNet, unfolded MP
& Large-scale, correlated, time-varying MIMO; model mismatch
& Small/well-conditioned systems; matched channel models
& Moderate-to-high inference burden; energy improves only after compression/accelerator support
& Re-training, latency, hardware mapping
\\

\textbf{MU-MIMO precoding}
& Wiener/MMSE, iterative nonlinear precoding
& Deep unfolding, learnable iterative optimization
& Dense multi-user settings; strong interference; dynamic channels
& Stable regimes where linear precoding is sufficient
& Moderate latency; energy cost rises with online adaptation and frequent updates
& Real-time adaptation, inference cost
\\

\textbf{Channel coding/decoding}
& LDPC BP/min-sum; polar SC/SCL/CA-SCL
& Neural BP, CNN/RNN/ Transformer decoders, RL/DRL code design
& Short block lengths; nonlinear/mismatched channels; correlated noise
& Long blocks; AWGN-like standard settings
& Low-to-moderate inference for compact/unfolded models, but high training energy for deep architectures
& Generalization, training data, complexity
\\

\textbf{Equalization}
& Linear equalization; iterative equalization--decoding
& DNN/CNN/RNN equalizers; joint learned pipelines
& Nonlinear distortion; strong ISI; hard-to-model channels
& When classical equalizers already preserve decoder assumptions
& Sequence-length-dependent latency; energy increases for long sequences and repeated retraining
& Sequence scaling, retraining overhead
\\

\textbf{NOMA / SCMA}
& DMPA; CSI-based detection; manual codebooks
& CNN/RNN detection; blind/image-based methods; D-SCMA
& Highly non-orthogonal access; CSI uncertainty
& Reliable-CSI cases with well-tuned classical detectors
& Moderate inference burden; energy overhead increases when detection and codebook learning are both learned
& CSI quality, codebook-learning complexity
\\

\textbf{Channel estimation}
& LS, LMMSE, compressed sensing
& LDAMP, DnCNN denoising, deep pilot design
& Noisy, fast-varying, structured channels
& Simple, well-modeled channels with low overhead needs
& Often edge-feasible at inference if compact, but training and denoising models can be energy-intensive
& Scalability, edge deployment cost
\\

\bottomrule
\end{tabular}
\end{adjustbox}
\vspace{-2mm}
\end{table*}
\rc{Overall, the evidence surveyed in this review suggests that AI-based methods are most valuable in complex, nonlinear, mismatched, or rapidly varying regimes, whereas classical methods remain highly competitive in well-modeled, standardized, and resource-constrained settings. In many cases, the most practical direction is therefore not full replacement, but a hybrid model-driven/data-driven design.}

\rc{The above sections focus on performance enhancement using AI. The increasing reliance on data-driven models also introduces critical security and privacy challenges, which are discussed in the next section.}
\begin{table*}[htbp!]
\centering
\caption{Comprehensive Overview of Key Contributions in Semantic Communication Systems}
\label{tabcomp3}
\resizebox{0.95\textwidth}{!}{%
\begin{tabular}{>{\raggedright\arraybackslash}p{2cm}>{\raggedright\arraybackslash}p{4cm}>{\raggedright\arraybackslash}p{3cm}>{\raggedright\arraybackslash}p{3cm}>{\raggedright\arraybackslash}p{3cm}>{\raggedright\arraybackslash}p{3cm}>{\raggedright\arraybackslash}p{3cm}}
\toprule
\textbf{Reference work} & \textbf{Key Contribution} & \textbf{Technique} & \textbf{Evaluation Metrics} & \textbf{Applications} & \textbf{Strengths} & \textbf{Limitations} \\ 
\midrule
J. Xu, et al. \cite{paper1WirelessImageTransmission2021} & Improved image transmission quality via adaptive resource allocation & ADJSCC with attention mechanisms & PSNR, LPIPS & Image transmission & Robust against channel mismatch & High computational demand \\ 
\midrule
P. Jiang, et al. \cite{paper2DeepSourceChannelCodingforSemantic2022} & Semantic transmission with feedback-based reliability & Semantic JSCC with HARQ & Sentence similarity, retransmission count & Text communication & High semantic accuracy & Limited dataset diversity \\ 
\midrule
J. Dai, et al. \cite{paper3NonlinearTransformSourceChannelCoding2022} & Improved semantic coding using nonlinear transforms & Nonlinear JSCC & Semantic similarity, bitrate efficiency & IoT, semantic comms & Efficient bitrate use & Scalability to complex data types \\ 
\midrule
M. Yang, et al. \cite{paper4MingyuYang-OFDMGuidedDeepJSCC2022} & Enhanced performance in multipath fading & OFDM with deep JSCC & BER, spectral efficiency & Wireless comms in fading environments & Robust multipath handling & Limited real-world testing \\ 
\midrule
W. Zhang, et al. \cite{paper5PredictiveAdaptiveDeepCoding2023} & Proactive coding via CSI prediction & Predictive semantic JSCC & PSNR, prediction accuracy & Image transmission & Low latency & Requires precise CSI feedback \\ 
\midrule
M. Kim, et al. \cite{paper6LearningEnd-to-EndChannelCoding2023} & Novel diffusion-based channel coding & End-to-end channel coding & BER, computational efficiency & Data transmission & High accuracy & Computational complexity \\ 
\midrule
J. Huang, et al. \cite{paper7JointTaskandData-OrientedSemantic2024} & Task and data-aware optimization & Deep separate source-channel coding & Task accuracy, resource use & IoT, task-based comms & Task-specific optimization & Complex system design \\ 
\midrule
V. Rana, et al. \cite{paper8ShortBlocklength2023} & Security-focused channel coding & DL for short blocklength codes & Error probability, security metrics & Secure communication & Highly secure and robust in adversarial conditions & Limited blocklength scenarios \\ 
\midrule
K. Yang, et al. \cite{paper9SwinJSCC2024} & Transformer-based semantic coding & SwinJSCC & PSNR, computational demand & Image transmission & High quality & High computation \\ 
\midrule
M. Merluzzi, et al. \cite{paper10Hexa-XProjectVision2023} & AI/ML for 6G communication and computation & Various ML techniques & Network throughput, energy efficiency & 6G, AI comms & Comprehensive 6G vision & Conceptual, no implementation \\ 
\midrule
Z. Weng, et al. \cite{paper11DeepLearningEnabledSemantic2023} & Speech-focused semantic communication & Semantic JSCC for speech & WER, PESQ & Speech transmission & Superior performance under low SNR & Limited real-world deployment \\ 
\midrule
Y. E. Sagduyu, et al. \cite{paper12Task-OrientedCommunications2023} & Task-oriented optimization with AI & DL and AI security & Task success rate & IoT, task-based comms & Security and task focus & Lacks scalability tests \\ 
\midrule
E. Erdemir, et al. \cite{paper13GenerativeJointSource-ChannelCoding2023} & Generative AI for image coding & GAN-based JSCC & Perceptual metrics, bitrate & Semantic image comms & High perceptual quality & High training cost \\ 
\midrule
R. Sangeetha, et al. \cite{paper14DeepLearning-BasedChannelEstimation2018} & Enhanced MIMO channel estimation & DL for channel estimation & Spectral efficiency, accuracy & Massive MIMO systems & High spectral efficiency & Limited scalability \\ 
\bottomrule
\end{tabular}%
}
\end{table*}
\begin{figure}[t]
    \centering
    \begin{tikzpicture}
    \begin{groupplot}[
        group style={
            group size=3 by 1,
            horizontal sep=1.7em
        },
        width=0.31\linewidth,
        height=4.7cm,
        xmin=0, xmax=20,
        xtick={0,5,10,15,20},
        xlabel={SNR (dB)},
        xlabel style={font=\footnotesize},
        ticklabel style={font=\scriptsize},
        label style={font=\footnotesize},
        grid=major,
        grid style={densely dotted, gray!35},
        axis line style={black},
        tick style={black},
        every axis plot/.append style={line width=1pt},
        every axis title/.style={
            at={(0.5,1.02)},
            anchor=south,
            font=\footnotesize
        },
        legend style={
            draw=none,
            fill=none,
            font=\footnotesize,
            legend columns=2,
            column sep=1em
        },
        legend to name=perflegend,
    ]

    \nextgroupplot[
        title={(a) PSNR (dB)},
        ymin=18, ymax=52,
        ytick={20,30,40,50}
    ]
    \addplot+[black, mark=* , mark size=1.8pt]
        coordinates {(0,24) (5,30) (10,36) (15,42) (20,48)};
    \addlegendentry{AI-based}
    \addplot+[gray!70!black, dashed, mark=square* , mark size=1.8pt]
        coordinates {(0,20) (5,26) (10,32) (15,38) (20,44)};
    \addlegendentry{Traditional}

    \nextgroupplot[
        title={(b) LPIPS ($\downarrow$)},
        ymin=0, ymax=0.55,
        ytick={0,0.1,0.2,0.3,0.4,0.5}
    ]
    \addplot+[black, mark=* , mark size=1.8pt]
        coordinates {(0,0.45) (5,0.35) (10,0.25) (15,0.15) (20,0.05)};
    \addplot+[gray!70!black, dashed, mark=square* , mark size=1.8pt]
        coordinates {(0,0.50) (5,0.40) (10,0.30) (15,0.20) (20,0.10)};

    \nextgroupplot[
        title={(c) BER ($\downarrow$)},
        ymin=0, ymax=16,
        ytick={0,4,8,12,16}
    ]
    \addplot+[black, mark=* , mark size=1.8pt]
        coordinates {(0,9.4) (5,7.2) (10,5) (15,2.8) (20,0.7)};
    \addplot+[gray!70!black, dashed, mark=square* , mark size=1.8pt]
        coordinates {(0,13.2) (5,11) (10,8.8) (15,6.3) (20,4)};

    \end{groupplot}

    \node at ($(group c1r1.north)!0.5!(group c3r1.north) + (0,0.7cm)$) {\ref{perflegend}};
    \end{tikzpicture}

    \caption{Illustrative performance \rc{trends} of AI-based and traditional channel coding schemes across SNR. Higher is better for PSNR, while lower is better for LPIPS and BER. Curves are \rc{adapted from \cite{paper3NonlinearTransformSourceChannelCoding2022}} and shown as indicative trends rather than directly comparable quantitative results.}
    \label{performance_comparison}
\end{figure}
%
\begin{figure}[t]
    \centering
    \begin{tikzpicture}
        \begin{axis}[
            width=0.88\linewidth,
            height=5.6cm,
            xmin=0.5, xmax=14.5,
            ymin=40, ymax=80,
            xlabel={Selected references},
            ylabel={Normalized BER reduction (\%)},
            xlabel style={font=\footnotesize, yshift=-0.2em},
            ylabel style={font=\footnotesize},
            xtick={1,...,14},
            xticklabels={
                {\cite{paper14DeepLearning-BasedChannelEstimation2018}},   
                {\cite{paper7JointTaskandData-OrientedSemantic2024}},      
                {\cite{paper9SwinJSCC2024}},                               
                {\cite{paper8ShortBlocklength2023}},                       
                {\cite{paper4MingyuYang-OFDMGuidedDeepJSCC2022}},          
                {\cite{paper1WirelessImageTransmission2021}},              
                {\cite{paper2DeepSourceChannelCodingforSemantic2022}},     
                {\cite{paper3NonlinearTransformSourceChannelCoding2022}},  
                {\cite{paper5PredictiveAdaptiveDeepCoding2023}},           
                {\cite{paper6LearningEnd-to-EndChannelCoding2023}},        
                {\cite{paper10Hexa-XProjectVision2023}},                   
                {\cite{paper11DeepLearningEnabledSemantic2023}},           
                {\cite{paper12Task-OrientedCommunications2023}},           
                {\cite{paper13GenerativeJointSource-ChannelCoding2023}}    
            },
            xticklabel style={
                rotate=90,
                anchor=east,
                font=\tiny
            },
            yticklabel style={font=\scriptsize},
            grid=major,
            ymajorgrids=true,
            xmajorgrids=false,
            grid style={densely dotted, gray!35},
            axis line style={black},
            tick style={black},
            enlarge x limits=0.02
        ]

        \addplot[
            ycomb,
            line width=0.8pt,
            gray!60,
            mark=none
        ] coordinates {
            (1,65.0)   
            (2,56.0)   
            (3,52.4)   
            (4,59.1)   
            (5,55.0)   
            (6,50.0)   
            (7,55.6)   
            (8,45.5)   
            (9,63.2)   
            (10,70.6)  
            (11,62.5)  
            (12,73.3)  
            (13,55.6)  
            (14,75.0)  
        };

        \addplot[
            only marks,
            mark=*,
            mark size=2.2pt,
            black
        ] coordinates {
            (1,65.0)
            (2,56.0)
            (3,52.4)
            (4,59.1)
            (5,55.0)
            (6,50.0)
            (7,55.6)
            (8,45.5)
            (9,63.2)
            (10,70.6)
            (11,62.5)
            (12,73.3)
            (13,55.6)
            (14,75.0)
        };

        \end{axis}
    \end{tikzpicture}
    \caption{Illustrative normalized BER reduction of AI-based methods relative to conventional baselines across selected 14 studied, synthesized from reported literature. \rc{Values are shown as qualitative trend indicators rather than directly comparable quantitative measurements.}}
    \label{fig:ber_reduction_lollipop}
\end{figure}
%
\begin{figure*}[htbp!]
\includegraphics[width=0.7\linewidth]{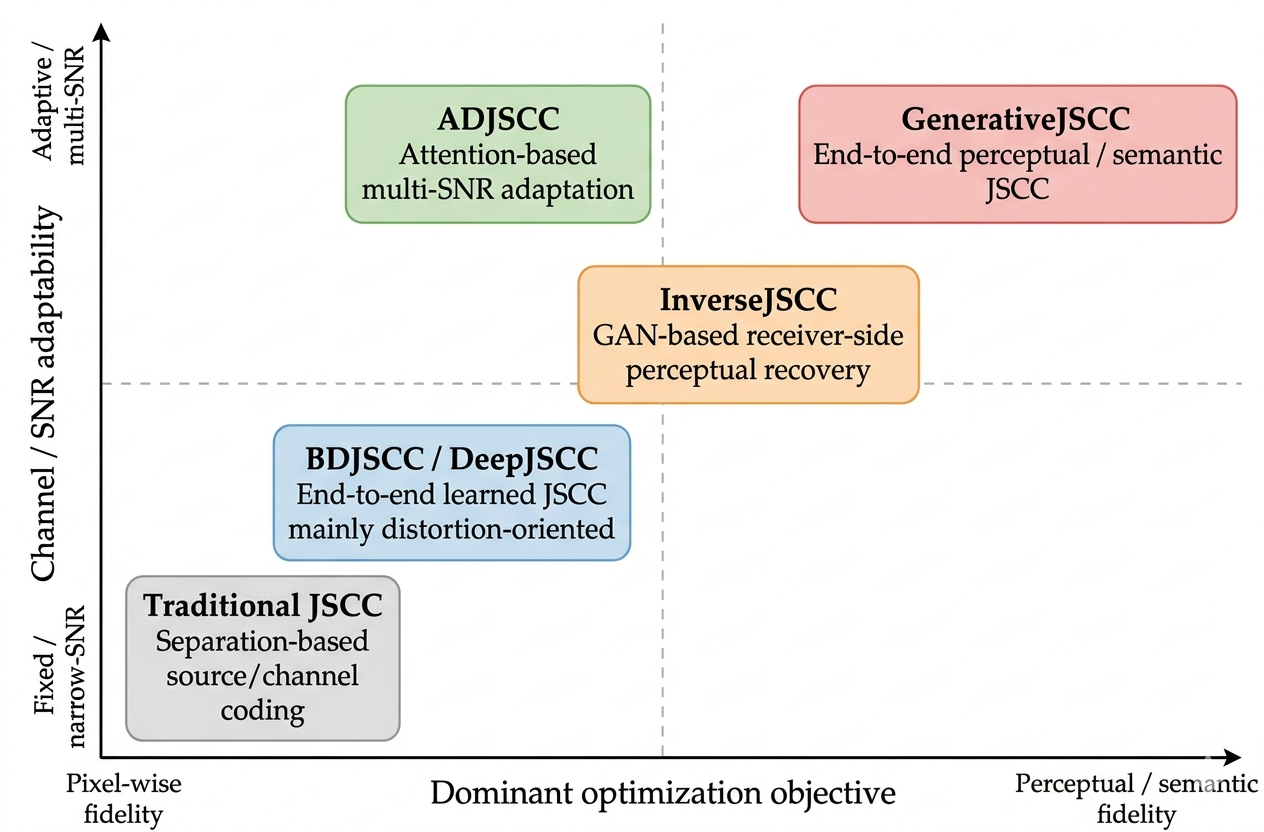}
\centering
\caption{Qualitative positioning of representative JSCC paradigms in wireless image transmission, synthesized from the methodological emphases and reported findings in \cite{paper1WirelessImageTransmission2021,paper13GenerativeJointSource-ChannelCoding2023}. The horizontal axis reflects the dominant optimization emphasis, from pixel-wise to perceptual/semantic fidelity, while the vertical axis reflects channel/SNR adaptability.}
\label{fig:jscc_positioning}
\end{figure*}

\section{Security and Privacy in AI-Driven 5G+ networks}\label{SecurityandPrivacy}
With the integration of AI into 5G/5G+ wireless networks, significant improvements in network management, performance, and resource allocation are expected \cite{zhangAIfor5G&B5G2020,LiEtAl_AIREV_2025_AISurvey_ChannelEstimation}. However, AI’s increasing role in telecommunications raises fresh security and privacy concerns \cite{Waqas_WirelessSecurity_AI_Review, xu2020channel}. These concerns become even more critical in 5G+ networks, where the transmission of massive amounts of sensitive data, complex network operations, and the broad diversity of connected devices heighten the risk. This section investigates the significance of security and privacy in AI-enhanced 5G/5G+ systems, the potential weaknesses in AI models, and the various strategies to safeguard these networks.

\subsection{Potential Vulnerabilities in AI Models Used in Telecommunications}
AI models in telecommunications, especially within 5G/5G+ networks, face several critical vulnerabilities, \rc{particularly when deployed at the physical layer for tasks such as channel estimation, decoding, and resource allocation \cite{LyYao_OJCOMS_2021_DL5GReview, nguyen2021application, paper4MingyuYang-OFDMGuidedDeepJSCC2022}}. A major concern is their susceptibility to adversarial attacks, where malicious inputs are crafted to mislead AI models \cite{ali2018secure}. \rc{Unlike traditional digital attacks, these perturbations can be embedded within wireless signals and propagated through the channel, making detection more challenging \cite{he2023adversarial}.} In 5G+ systems, such attacks could disrupt key functions like resource allocation, scheduling, and beamforming, potentially leading to network failures, congestion, or sensitive data breaches \cite{ali2017stabilizing}. \rc{In AI-driven channel coding and decoding systems, such attacks can significantly degrade BER performance, especially under low SNR conditions or model mismatch scenarios \cite{ly2021review, garg2021security}.} Another threat is data poisoning, where attackers inject malicious data during training. \rc{In distributed or federated training environments, this can lead to biased or unreliable decoding models, ultimately compromising communication reliability \cite{LyYao_OJCOMS_2021_DL5GReview, Waqas_WirelessSecurity_AI_Review}.} Model inversion attacks pose significant privacy risks by allowing attackers to infer sensitive information (e.g., user location or preferences) from model outputs \cite{BagheriTowardConnected2021}. Moreover, due to model transferability, an attacker can replicate or adapt a trained model to exploit other systems, even under different datasets or environments \cite{fereidooni2021safelearn}. \rc{Furthermore, the transferability of adversarial attacks allows malicious perturbations crafted in one environment to remain effective across different channel conditions \cite{he2023adversarial}.} These vulnerabilities underscore the need for robust AI security mechanisms in future telecom infrastructures.
\subsection{Approaches to Ensuring Privacy in AI-based Communication Systems}
To secure AI-driven 5G+ networks, several advanced techniques are being developed. Adversarial ML \cite{he2023adversarial} strengthens AI models against manipulation by exposing them to adversarial examples during training, increasing their resilience to attacks, especially in critical functions like traffic management. Secure FL \cite{fereidooni2021safelearn} enables decentralized AI training by keeping raw data on local devices and sharing only model updates. \rc{This approach is particularly relevant for distributed wireless systems, where collaborative training of channel coding and decoding models can be achieved without exposing sensitive user data \cite{chen2020wireless}.}  \rc{However, FL introduces new vulnerabilities, including gradient leakage and model inversion during aggregation \cite{liu2020federated}. To mitigate these risks, secure aggregation protocols and encryption-based techniques are employed to protect model updates during transmission \cite{fereidooni2021safelearn}.} Differential chaotic privacy \cite{ali2023chaos, 11062788} further protects sensitive information by adding noise to AI outputs, preventing the disclosure of individual data points. \rc{Such privacy-preserving mechanisms must carefully balance data protection with communication overhead and model accuracy.} This is especially important in applications where privacy is paramount, such as healthcare and finance. 
However, certified robustness for neural decoders and AI-based resource allocators remains immature, especially for safety-critical applications such as URLLC, autonomous driving, and industrial control, where formal reliability guarantees are required under adversarial or distribution-shift conditions.
\subsection{Through Quantum Communication}
Quantum communication, including Quantum Key Distribution (QKD) and quantum encryption, is expected to become increasingly important in 5G+ and future wireless networks (as targeted in releases 19 and 20) \cite{imran2024quantum}. \rc{However, its relevance to current AI-driven wireless communication systems remains limited due to significant practical constraints, including specialized hardware requirements, scalability challenges, and integration complexity \cite{imran2024quantum}.} \rc{As a result, quantum communication is better viewed as a long-term research direction rather than an immediate solution for securing AI-based 5G/5G+ networks.}
\subsection{Advantages and Limitations of AI-based Solutions in 5G/5G+}
\begin{figure*}[htbp!]
\includegraphics[width=0.99\linewidth]{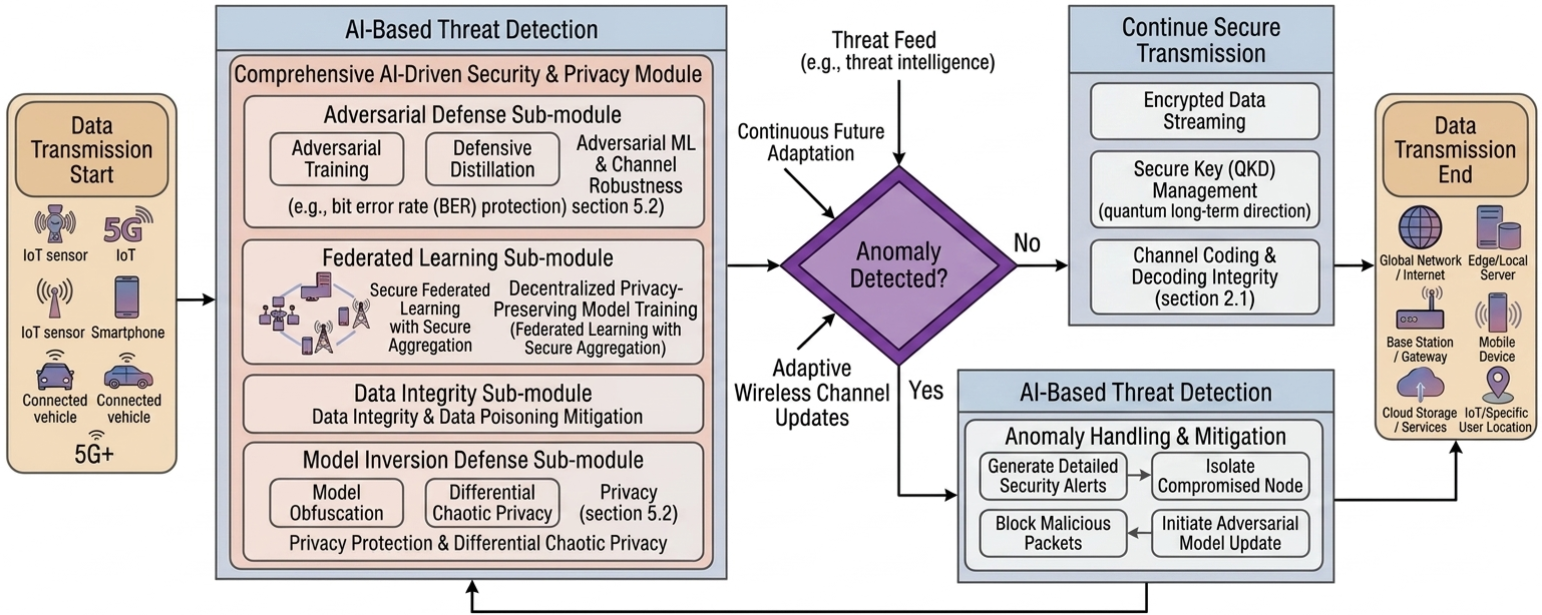}
\centering
\caption{\rc{Conceptual illustration of AI-assisted threat detection in wireless data transmission.}}
\label{Fig13}
\end{figure*}

AI-based solutions offer several advantages for 5G/5G+ networks, notably real-time automation and optimization. By learning from network data, AI algorithms enable dynamic decisions on resource allocation \cite{ly2021review}, routing \cite{HuMLbasedAdaptiverouting2010}, and scheduling \cite{LiDecAge2024, yasiraliinterless}, thus ensuring efficient operation. They also facilitate proactive management through anomaly detection and failure prediction, reducing downtime. Moreover, AI supports scalable data handling, making it ideal for dense 5G+ environments with massive device connectivity. However, limitations remain. AI model training and inference incur high computational overhead, especially for real-time or edge-deployed systems \cite{garg2021security}. Model complexity also increases with network scale, complicating deployment and maintenance \cite{9447977, freire2023reducing}. Additionally, reliance on large datasets raises privacy concerns due to the potential exposure of sensitive user data. \rc{Furthermore, AI-based communication models require continuous adaptation to dynamic wireless environments \cite{ dai2023toward, lyu2018performance}. Changes in channel statistics, mobility patterns, or interference conditions may degrade model performance, necessitating retraining or fine-tuning, which introduces additional computational and communication overhead.} Fig.~\ref{Fig13} illustrates \rc{a simplified conceptual framework of} AI-driven security framework. Starting at ``Data Transmission Start," AI analyzes traffic for threats. If anomalies are detected, the system triggers alerts and blocks malicious data; otherwise, secure transmission continues uninterrupted.  \rc{However, real-world implementations require more advanced mechanisms, including robust adversarial defenses, secure model updates, and integration with physical-layer security techniques to ensure reliable operation.}
\section{Future Research Directions in AI for 5G/5G+}\label{FutureResearchDirections}
AI integration in 5G+ networks presents significant potential, alongside challenges and opportunities for future exploration. Research efforts focus on enhancing methodologies, optimizing performance, and addressing outstanding issues to enable sustainable and effective AI deployment in next-generation networks.
\subsection{Open Research Challenges}
Despite promising advancements, several open challenges must be addressed to fully realize the potential of AI in 5G/5G+ networks.
\subsubsection{AI Model Optimization}
A key challenge lies in balancing AI model complexity with real-time and resource-constrained environments. DNNs, while powerful, are computationally intensive and unsuitable for edge deployment \cite{lyu2018performance}. Efforts are focused on lightweight architectures using pruning, quantization, and knowledge distillation to reduce model size and latency. However, scalability remains difficult due to heterogeneous 5G+ topologies, dynamic channels, and diverse service types (e.g., URLLC, mIoT). Future research must also address cross-device interoperability and energy efficiency. \rc{A key open research question is how to design neural decoders that can achieve near-optimal performance (e.g., within 0.5--1 dB of BP decoding) while reducing computational complexity by at least 50\% for real-time edge deployment. Future research should also explore hybrid model-driven and data-driven architectures that incorporate domain knowledge to reduce training overhead and improve generalization. Additionally, comparative evaluation of compression techniques such as pruning versus quantization under varying SNR conditions is needed to identify optimal trade-offs between performance and efficiency.}
\subsubsection{Data Privacy}
AI in 5G/5G+ often relies on user-sensitive data such as location or behavior \cite{BagheriTowardConnected2021}, raising privacy concerns. Techniques like FL and differential privacy aim to keep data local \cite{ELRAJAB2024110294}. However, maintaining model accuracy while preserving privacy remains challenging. Further work is needed to balance privacy protection with the benefits of large-scale, diverse datasets. \rc{An important research direction is to develop privacy-preserving training mechanisms that maintain performance within 5--10\% of centralized training accuracy while ensuring strong privacy guarantees. This includes investigating trade-offs between differential privacy noise levels and decoding accuracy. Furthermore, comparative studies between FL and centralized approaches in dynamic wireless environments are required to quantify communication overhead, convergence speed, and robustness to adversarial participants.}
\subsubsection{AI Fairness}
Ensuring fairness in AI-driven decision-making is critical as 5G+ networks expand globally. Bias in training data can lead to unequal service quality or resource distribution across demographics or regions \cite{EiEfficientResourceAllocation2022}. Solutions such as fairness constraints and explainable AI are under development \cite{ly2021review}, but standardized approaches are still lacking. Ensuring equitable outcomes remains an ongoing research priority.
\rc{Future research should define measurable fairness metrics specific to wireless systems, such as equitable resource allocation across users with varying channel conditions. A key challenge is to ensure that fairness constraints do not degrade overall system throughput beyond acceptable limits (e.g., less than 10\% reduction). Comparative analysis of fairness-aware optimization algorithms versus conventional resource allocation methods is necessary to identify scalable and practical solutions.}
\subsubsection{Real-Time Decision-Making}
Applications like autonomous driving, healthcare, and industrial automation demand AI decisions within milliseconds \cite{ELRAJAB2024110294, AI-BasedAutonomousControl2020}. Meeting these strict latency constraints in edge environments is difficult due to limited compute and bandwidth. Research is needed to design fast, efficient, and adaptive AI models capable of operating under uncertainty and varying network conditions, such as congestion and mobility.
\rc{A critical research question is how to design AI models capable of making decisions within strict latency bounds (e.g., sub-millisecond inference for URLLC applications) while maintaining high reliability. Future work should explore lightweight architectures and hardware-aware model design to achieve inference times comparable to traditional algorithms. Comparative benchmarking of AI-based versus classical approaches in terms of latency, throughput, and reliability is essential to assess practical feasibility.}
\rc{
\subsubsection{AI-Driven Channel Coding: Open Research Questions}
Despite recent progress, several fundamental questions remain unresolved in AI-based channel coding. A key challenge is identifying the precise conditions under which DL-based decoders outperform classical schemes such as LDPC and polar codes. This requires systematic evaluation across varying block lengths, channel models, and SNR regimes. Another important direction is the development of generalizable neural decoders that can maintain performance across diverse channel conditions without requiring frequent retraining. Current models often suffer from performance degradation under distribution shifts, highlighting the need for robust and adaptive architectures. Furthermore, there is a need to establish standardized benchmarks for comparing AI-based and classical coding techniques. These benchmarks should include metrics such as BER, latency, energy consumption, and computational complexity under consistent evaluation settings. Finally, future research should investigate hybrid decoding strategies that combine model-based and data-driven approaches, aiming to achieve near-optimal decoding performance with reduced complexity.}
\rc{
\subsection{Practical Deployment Challenges and System-Level Trade-offs}
While AI-based techniques have demonstrated performance improvements in 5G+ systems, their practical deployment introduces several system-level challenges that must be carefully addressed.
\subsubsection{Model Compression and Performance-Complexity Trade-offs}
Model compression techniques such as pruning, quantization, and knowledge distillation are commonly employed to reduce the computational and memory overhead of DL models \cite{gupta2022compression, wang2023versatile}. Pruning can significantly reduce the number of parameters, often by 50–90\% with minimal degradation in performance when applied carefully \cite{mashhadi2021pruning}. Similarly, quantization techniques (e.g., 8-bit or even 4-bit representations) can reduce memory footprint and accelerate inference, typically incurring only marginal performance loss (e.g., <1 dB in BER performance in many cases) \cite{keller202395}. However, aggressive compression may lead to noticeable degradation, especially in low-SNR regimes where model precision is critical. Knowledge distillation offers a compromise by transferring knowledge from large models to smaller ones, achieving reduced complexity while retaining most of the performance benefits \cite{fang2026knowledge}. These trade-offs highlight the need for careful optimization depending on system constraints such as available hardware resources and latency requirements.
\subsubsection{Hardware Implementation Considerations}
The deployment of AI-based communication models on real-world hardware platforms remains a significant challenge \cite{moghaddam2023statistical}. Unlike classical decoding algorithms, which are highly optimized for hardware implementation (e.g., ASICs and FPGAs), DNNs often require specialized accelerators, such as GPUs or AI-specific hardware (e.g., TPUs and NPUs) \cite{kamath2025asic}. Mapping neural decoders to hardware introduces challenges related to memory access patterns, parallelism, and power consumption. Additionally, irregular network architectures may limit efficient hardware utilization \cite{Aliieeeaccess2024}. Recent advances in edge AI hardware and neuromorphic computing offer promising solutions; however, standardized implementation frameworks for communication systems remain under development. Concurrently, emerging research highlights the importance of hardware–software co-design for AI-based decoders, wherein neural architectures are jointly optimized alongside FPGA/ASIC implementations to satisfy latency, energy efficiency, and real-time deployment constraints \cite{1bitMassiveMUMIMO2017,kamath2025asic}. Moreover, hardware reliability is particularly important in specialized communication platforms such as satellite-enabled 5G systems, where memory robustness and soft-error resilience can directly affect practical deployment, thereby motivating reliability-aware circuit and architecture design alongside algorithmic optimization.
\subsubsection{Latency and Real-Time Constraints}
Latency is a critical performance metric in 5G+ systems, particularly for URLLC scenarios. While classical decoding algorithms typically have predictable and optimized latency, neural network inference introduces additional computational overhead \cite{jiang2020learn}. Inference latency depends on factors such as network depth, parameter count, and hardware acceleration. In many cases, AI-based decoders may exhibit higher latency than traditional methods unless optimized through model compression and parallel processing. Therefore, achieving real-time performance requires careful co-design of algorithms and hardware.
\subsubsection{Energy Efficiency Considerations}
Energy consumption is another key concern, particularly for mobile and edge devices. AI-based models generally consume more energy during inference due to increased computational complexity and memory access requirements \cite{keller202395}. However, energy efficiency can be improved through model compression, hardware acceleration, and efficient scheduling. In contrast, classical communication algorithms are typically more energy-efficient due to their structured and iterative nature. As a result, the adoption of AI-based techniques must be justified by significant performance gains that outweigh their energy costs.
\subsubsection{Adaptability and Retraining Overhead}
One of the fundamental challenges in deploying AI-based communication systems is their dependence on training data \cite{lv2023deep}. When channel statistics change due to mobility, environmental variation, or interference, pre-trained models may no longer perform optimally \cite{Waqas_WirelessSecurity_AI_Review}. Adapting to such changes often requires retraining or fine-tuning, which introduces additional computational overhead and latency. Online learning and transfer learning approaches have been proposed to mitigate this issue, but they remain an active area of research. This limitation contrasts with classical methods, which are typically more robust to changing conditions due to their model-based design.}
\subsection{Future Directions of AI-Driven Learning in Wireless Networks}
AI-driven learning in 5G/5G+ networks offers promising avenues but also presents technical challenges. While traditional methods remain useful for simpler tasks due to their lower complexity, AI excels in complex, hard-to-model scenarios. Model-driven learning leverages domain knowledge for efficient training, whereas data-driven approaches capture intricate patterns but require higher computational resources. A hybrid strategy may offer a practical tradeoff. Ensuring data quality accuracy, integrity, and fairness is essential, requiring robust preprocessing and handling of biases. Barriers like data silos and a lack of standardization impede deployment. Secure, unified data management protocols are necessary to facilitate broader AI integration. Training strategy also matters: offline learning enables controlled optimization, while online learning offers adaptability at the cost of higher resource demands. Most current AI solutions target individual modules (e.g., channel estimation or detection). Joint learning across modules could improve performance, but it adds complexity. Similarly, cross-layer learning applying AI across multiple network layers can boost adaptability and efficiency beyond the physical layer. Implementation choices also influence performance. Software-based AI offers flexibility but is resource-intensive. In contrast, hardware-based solutions (e.g., FPGAs and ASICs) offer lower power consumption and faster inference but lack flexibility. Future work should prioritize hybrid, hardware-efficient AI architectures to achieve scalable and secure network optimization.

\section{Conclusion}\label{Conclusion}
\rc{This survey examined AI-driven channel coding and coding-aware resource optimization for wireless networks, with emphasis on MIMO detection and precoding, channel decoding, equalization, NOMA/SCMA, channel estimation, and security. A central insight from the reviewed literature is that AI-based methods are most effective in complex wireless regimes, such as nonlinear, mismatched, highly dynamic, or interference-limited environments, where model-based methods become inaccurate or overly rigid. Classical model-based methods remain highly competitive in standardized settings with well-characterized channels and tight resource constraints, particularly for long-block coding and low-complexity deployment. Thus, the evidence indicates that AI is not viewed as a universal replacement for conventional communication algorithms; instead, it points to hybrid model-driven/data-driven architectures as the most practical path forward. For channel coding specifically, neural decoders show promise under short block lengths, correlated noise, and model mismatch, but their gains are often inconsistent across studies and remain limited by generalization, training cost, latency, and hardware efficiency. Based on the surveyed literature, the most important future directions are: standardized benchmarking across BER, latency, energy, and complexity; robust neural decoders that can generalize across channel variations without frequent retraining; lightweight and compressed models for real-time edge deployment; hardware-software co-design; and privacy-preserving, attack-resilient distributed learning. These directions are essential for translating promising AI-based communication methods into reliable and scalable wireless systems.}

\section*{CRediT Authorship Contribution Statement}

\textbf{Yasir Ali}: Writing – review \& editing, Writing – original draft, Visualization, Validation, Software, Methodology, Investigation, Formal analysis, Conceptualization.
\textbf{Tayyab Manzoor}: Writing – review \& editing, Project administration, Conceptualization. \textbf{Huan Yang}: Writing – review \& editing, Visualization, Validation.  \textbf{Chenhang Yan}: Writing - review \& Editing. \textbf{Yuanqing Xia}: Writing – review \& editing, Supervision, Visualization, Validation, Formal analysis, Funding acquisition.

\section*{Declaration of Competing Interest}

The authors declare that they have no known competing financial interests or personal relationships that could have appeared to influence the work reported in this paper.

\section*{Acknowledgments}
This work was supported in part by the Beijing Natural Science Foundation Haidian Original Innovation Joint Fund Project under Grant L252035, in part by the National Natural Science Foundation of China under Grant U25A20460, in part by the National Natural Science Foundation of China for International Young Scientists under Grant W2533176, in part by the Beijing Natural Science Foundation under Grant No. IS25064, in part by the Joint Fund Project of Shandong Provincial Natural Science Foundation under Grant
ZR2025LZH001, in part by the Henan Provincial Talent Program under Grant 264000510006 and in part by the Henan Postdoctoral Foundation under Grant HN2026058
\section*{Data availability}
Data will be made available on request.

\bibliographystyle{elsarticle-num-names} 
\bibliography{bib}
\end{document}